\documentclass[twocolumn]{bmcart}

\usepackage{amsthm,amsmath}
\usepackage{threeparttable}
\usepackage{adjustbox}
\usepackage{multirow}

\usepackage[numbers,sort&compress]{natbib}
\usepackage{graphicx}

\usepackage{subfigure}
\usepackage{tabularx} 
\usepackage{caption}

\usepackage{threeparttable} 
\usepackage[figuresright]{rotating}
\usepackage{endnotes,amsmath}
\usepackage{makecell}
\usepackage{booktabs}
\usepackage{array}
\usepackage{bm}
\usepackage{url} 
\usepackage[figuresright]{rotating}
\usepackage[utf8]{inputenc} 
\usepackage[ruled,linesnumbered]{algorithm2e}

\startlocaldefs
\endlocaldefs

\begin{document}

\begin{frontmatter}

\begin{fmbox}
\dochead{Research}


\title{Towards the Universal Defense for Query-Based Audio Adversarial Attacks}


\author[
    addressref={aff1,aff2},                   
   email={guo\_feng@mail.sdu.edu.cn }
]{\inits{GF}\fnm{Feng} \snm{Guo}}
\author[
  addressref={aff1,aff2},
  email={sun\_zheng@mail.sdu.edu.cn }
]{\inits{SZ}\fnm{Zheng} \snm{Sun}}
\author[
  addressref={aff1,aff2},
  corref={aff1,aff2},    
  email={chenyuxuan@sdu.edu.cn}
]{\inits{CYX}\fnm{Yuxuan} \snm{Chen}}
\author[
  addressref={aff1,aff2},
  email={julei@sdu.edu.cn}
]{\inits{JL}\fnm{Lei} \snm{Ju}}

\address[id=aff1]{
  \orgdiv{School of Cyber Science and Technology},             
  \orgname{Shandong University},          
  \city{Qingdao},                              
  \cny{China}                                    
}
\address[id=aff2]{
  \orgdiv{Quancheng Laboratory},         
  \orgname{Shandong University},          
  \city{Jinan},                              
  \cny{China}                                    
}



\end{fmbox}


\begin{abstractbox}

\begin{abstract} 
Recently, studies show that deep learning-based automatic speech recognition (ASR) systems are vulnerable to adversarial examples (AEs), which add a small amount of noise to the original audio examples. These AE attacks pose new challenges to deep learning security and have raised significant concerns about deploying ASR systems and devices. The existing defense methods are either limited in application or only defend on results, but not on process. In this work, we propose a novel method to infer the adversary intent and discover audio adversarial examples based on the AEs generation process. The insight of this method is based on the observation: many existing audio AE attacks utilize query-based methods, which means the adversary must send continuous and similar queries to target ASR models during the audio AE generation process. Inspired by this observation, 
We propose a memory mechanism by adopting audio fingerprint technology to analyze the similarity  of the current query with a certain length of memory query.
Thus, we can identify when a sequence of queries appears to be suspectable to generate audio AEs. Through extensive evaluation on four state-of-the-art audio AE attacks, we demonstrate that on average our defense identify the adversary's intent with over $90\%$ accuracy. With careful regard for robustness evaluations, we also analyze our proposed defense and its strength to withstand two adaptive attacks. Finally, our scheme is available out-of-the-box and directly compatible with any ensemble of ASR defense models to uncover audio AE attacks effectively without model retraining.
\end{abstract}




\end{abstractbox}
%

\end{frontmatter}



\section*{Introduction}\label{introduction}
Benefiting from the application of deep learning, the field of speech recognition has also been widely developed. However, deep learning-based automatic speech recognition (ASR) systems are shown to be vulnerable to audio adversarial examples (AEs), which add tiny perturbations on benign audio clips to fool the deep neural network model. Thus, how to secure ASR systems to prevent AE attacks remains a critical question.

Multiple mechanisms have been proposed to defend against audio AEs on ASR. Some methods mainly rely on signal processing skills such as smoothing, downsampling, reconstruction, and so on \cite{10,11,15,16}. These methods can destroy the adversarial components of AE to a certain extent, and prevent them from reaching the preset target to reduce their impact on ASR. But it also destroys the benign sample and works for defense against unknown attacks. There are some works that train an additional DNN network as a prior part of ASR \cite{18,19,20}. However, those defense methods depend heavily on the algorithms for generating AEs, the generalization capability is the key that limits the ability of defense, and the model will be difficult to discriminate the adversarial samples without participating in the training. In addition, the existing defense methods against audio adversarial examples focus on the generation results of AEs, without on the process.

We reinvestigate and rethink the process of generating the adversarial examples, trying to locate the ``specific'' features in this process. We also scrutinize the current state-of-the-art attacks, including white-box attacks \cite{4,7,34}, black-box attacks \cite{multi-objective,API65,API66} and transfer attacks \cite{transferable-black-box-attacks,transfer60,transfer61}. 
We note that the perturbation of the AEs in some attacks is quite light, and the distance between them and the benign examples is small without a particularly significant difference. So it is difficult to identify whether a single input is an AE. We often ignore the process of AE generation and only pay attention to the results.  
How to utilize this discarded information. Yet, except for some attacks that directly generate AEs, the majority need to keep visiting the target model to adjust the AE, essentially stealing key information (e.g., gradients) from the model. In this case, the adversary needs to send massive and similar queries to the target model in a period, which likely exposure her adversarial behavior. Therefore, according to this feature, we do not try to discover individual inputs, rather we focus on the relationship between the inputs to recognize the attack.


In this work, we propose a universal and lightweight defense framework to infer the adversarial behavior by memory mechanism. 
The basic idea of our framework is that generating adversarial examples and the query to ASR models is continuous and correlated before and after. In contrast, a regular query is independent of others. 
We consider some history inputs of a certain length as a piece of memory, analyze the correlation between a new input and the memory, and mark the input as adversarial if the correlation crosses a certain threshold. 
We use the similarity of the audio fingerprint to estimate the correlation of the input. 
The insensitivity of the audio fingerprint to noise is an attractive trait. Meanwhile, since its simplicity, it is hard for the adversary to be aware of the use of defensive models.
Furthermore, motivated by the similarity matrix  for recommender systems,
In this way, we can efficiently and quickly verify that the input query sound is adversarial or benign.
We employ a non-neural network defense architecture and are not able to optimize the defense model in a similar way to a neural network, so an attacker may not be able to attack the defense model from that perspective.
This strategy efficiently identifies the existing state-of-the-art adversarial sample attacks. The robust average uncovering success rates ($DSR$) are all above $90\%$. Also, our proposed framework can be easily combined with any other existing defense methods. 

Finally, we study some adaptive attacks. 
We designed experiments with random noise attacks, which disturbed audio fingerprint feature extraction. For noise adaptive attacks, we observed that the modest level of random noise instead results in better performance to our defense system and we build a more robust defense system. In addition, we tested the potential role of different ``\textit{fake query}" ratios $ {p_{fake}}$ on the results.
We conducted experiments on both types of adaptive attacks and proved that our defense framework remains robust under the damage.

The main contributions of this work are three-fold: 

\quad \noindent$\bullet$ We propose a new defense mechanism for adversarial audio attacks by analyzing the correlation between input with memory. 
This is the first proposed defense framework based on the AEs generation process for the ASR.
The robust average uncovering success rates are all beyond $90\%$ for  existing attacks and we first evaluated the music-based AEs.

\quad \noindent$\bullet$ We demonstrate the robustness of our defense framework toward adaptive attacks. We found that the adaptive attack methods of fingerprint extraction damage and the ``\textit{fake query}'' are unable to evade our defense, and our defense strategy is still effective. 
We build a more robust defense system through the combination of a moderate level of random noise.

\quad \noindent$\bullet$ We designed a music-carrier dataset that can be used to produce audio adversarial examples, which also establish a foundation for future research on attacks and defenses based on music-carrier. 
And we release the source code for our defense and datasets at: \url{https://github.com/xxxx.}



\section*{Background and Related Work}\label{back}
\textbf{Adversarial Examples (AEs)}.
Adversarial attacks originate from images and quickly develop, with much relevant research. Many works achieve successful attacks on image classifiers by the computed gradient and these attacks are relatively convenient to implement \cite{2,49,50,52}. Some work explores transfer attacks from white-box to black-box models but needs a lot of access to the target model \cite{55,transfer60,transfer61}.
This provides a good reference for adversarial studies on audio. 
 One may inquire about the reasons for the existence of adversarial examples. According to several works \cite{45,46,47,48}, they think that adversarial examples are not a network drawback but a feature. The network attempts to learn ``all'' the beneficial features during the training process, whereas humans are naturally inclined to ignore some features.
 When an adversary attacks the model via manipulation of such features, it leads to a rapid decrease in the accuracy of the model, whereas the accuracy of humans is immune.
 Thus our concern is not to remove the AEs and it fails to do so, instead, we should avoid the risk of the AEs to the model.

\textbf{Audio Adversarial Attacks to ASR}.
A similar situation exists in the ASR. Typically, a state-of-the-art ASR model is susceptible to deception by malicious AEs, which has evolved from a single-word attack to an attack on the entire sentence. 
Some state-of-the-art models were successfully attacked, \cite{4} used CTC-loss to compute gradients to achieve an attack on DeepSpeech; CommanderSong \cite{7} used pdf-id to design a loss function to implement attack base on Kaldi \footnote{\url{http://kaldi-asr.org.}}; \cite{5}  implemented an attack on Lingvo \footnote{\url{https://github.com/tensorflow/lingvo}} with psychological masking. For black-box attacks, the gradient is incomputable. However, \cite{6} successfully attacked the DeepSpeech black-box model with a genetic algorithm; \cite{9} successfully attacked four commercial speech API services (Google Cloud Speech-to-Text, Microsoft Bing Speech Service, IBM Speech to Text, and Amazon Transcribe); \cite{15} successfully attacked the speech recognition API interfaces of iFLYTEK and Ali with the co-evolutionary algorithm.
Besides, already there are attacks that can be launched in the physical world. In order to enhance the robustness of physical attacks, in \cite{6,8,9}, the authors added the Gaussian white noise to AEs and the evaluation results show that this strategy enhances the physical robustness of the AEs. Although they do not require a specific noise model, they may rely on the playback device and the experimental environment.
These attacks inevitably require a massive amount of queries to models, and query-based attacks are becoming worse with time. In this article, our main object of our article will be focused on recognizing such attacks before they succeed and defending against query-based adversarial attacks.

\textbf{Defense against Audio Adversarial Attacks}. 
The majority of proposed methods of defense against audio adversarial attacks are removing or ruining the adversarial component by the technical tool of signal processing. Paper \cite{10} proposed random smoothing to mask the disturbing adversarial component. \cite{11} proposed WaveGAN vocoder to reconstruct the waveform to eliminate the disturbing domain. \cite{12} used label smoothing, \cite{13,14} squeezed the audio, \cite{15} is the down-sampling method and \cite{16} added distorted signals. These works of defense are concerned with removing or ruining the perturbation component. Those approaches have both advantages and disadvantages, as it breaks the adversarial behavior of AEs while also causing a lot of damage to examples of benign queries. Deficiency of hard evidence for the difference between AEs and benign examples. Some people suggested applying sub-models to preclude some attacks \cite{17,18}. The literature \cite{19,20,21} applies extra neural networks to check adversarial examples to protect the ASR model. But they can only restrain some existing attacks, which are impotent to uncertain attacks. The applications are limited due to the sub-models bulky. Some methods based on state detection of images \cite{26,advmind} also provide some guidance for the audio adversarial attacks. Although these defensive works are available for certain types of attacks, it is a deficiency that the evaluation of adaptive attacks is incomplete or oversimplified. No integral architecture is available for combination with other methods. We work mainly on building a lightweight framework that can be easily combined with other defense methods.

\begin{figure}[htbp]
\centering
\includegraphics[width=0.45\textwidth]{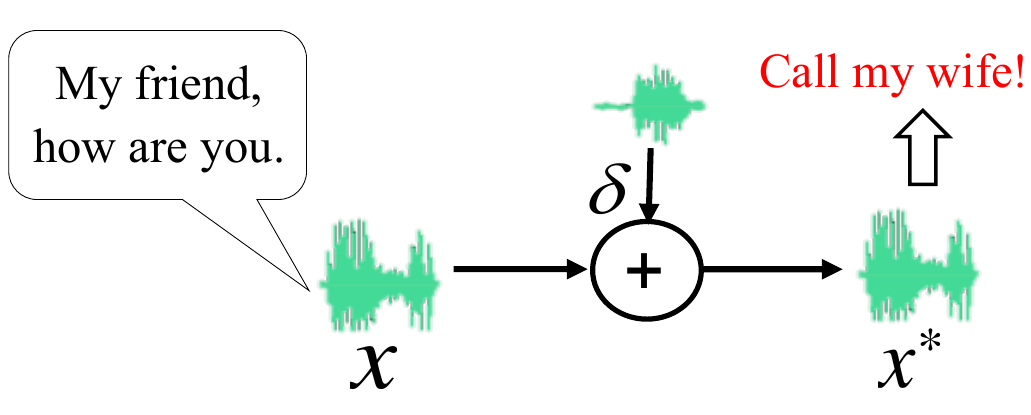}
\caption{\label{fig:1}The correct transcription of  $x$ is ``My friend, how are you'', and the adversary's purpose is to add a careful perturbation ``$\delta$'' to $x$ and then make it become ${x^{\rm{*}}}$ that can be transcribed as the target of ``Call my wife''.}
\end{figure}

\textbf{Problem Setup}.
Hereafter, we concentrate on adversarial tasks. In a setup like this, the DNN network is represented as $f$, and $f:\boldsymbol{X}\to \boldsymbol{C}$ represents the given input $x(x\in \boldsymbol{X})$ is mapped to one of a set of classes $\boldsymbol{C}$, where $f(x)=c\in \boldsymbol{C}$. The DNN model is vulnerable to adversarial input attacks, which forces the DNN model to misjudge. Attacks on DNNs can be classified as targeted and untargeted. Here, we will focus on the setting of targeted attacks. Specifically, adversarial examples ${x^*}$ are normally generated by slightly modifying $x$ and ${x^*} = x + \delta $. The solve of $\delta$ can be converted to a min-optimization problem, i.e., $arg$ $min$  $\mathcal L(f(x + \delta ),c^*)$. The adversary's goal is to force $f$ to misclassify ${x^*}$ as the target ${c^*}$, i.e., $f({x^*}) = {c^*} , {c^*} \ne c $. To ensure that ${x^*}$ is acoustically similar to $x$, the perturbation needs to be restricted to a limited range $g({x^*}-x) \le \varepsilon $, where the $g$ is a measure function of the auditory difference. The attack process is shown in Fig. \ref{fig:1}. 

\section*{Defense against Query-Based  Audio Adversarial Attacks}\label{ approaches}
\begin{figure}[htp]
\centering
\includegraphics[width=0.45\textwidth]{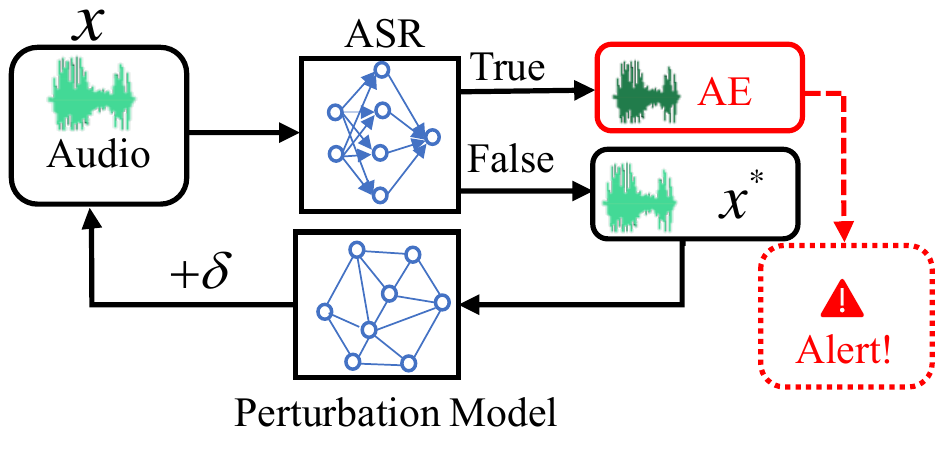}
\caption{\label{fig:2}Query-based attack: setting a target, for the first time ${x^{\rm{*}}}{\rm{ = }}x$, if $x^*$ can be transcribed as a target, the AE is true, else false,  adjust the $\delta$ carefully, and perform the next query. Repeat this process until  $x^*$ can be transcribed as the target.}
\end{figure}

A successful audio AE requires a specified carrier (the carrier can be music or dialogue) undergoing several iterations and queries. The process of AE generation is continuous. Every time, the adversary needs to produce a small disturbance $\delta$ to repeatedly adjust ${x^*}$. When crossing the decision boundary, a successful AE is done and the whole process is depicted in Fig. \ref{fig:2}. Our defense is motivated by the process nature of query-based attacks. We can examine the query-to-memory relationship to determine if queries are intended to generate an AE, which is the process-based defense approach. To calculate the correlation $C$ of the new query about the memory, we used the similarity $F$ of the audio fingerprint to estimate the correlation, i.e $C ({q_{{\rm{memory }}}} , {q_{{\rm{new }}}}) \approx F ({q_{{\rm{memory }}}} , {q_{{\rm{new }}}})$. For each query, audio has unique fingerprint information. The audio fingerprint is robust to noise and adapts to a noisy environment. Moreover, it can prevent audio splicing attack \cite{finger}. According to the obtained fingerprints, we can figure out the similarity between the input query and the memory, which provides the foundation for our determination.

\textbf{Defense Architecture}

\begin{figure}[htb]
\centering
\includegraphics[width=0.45\textwidth]{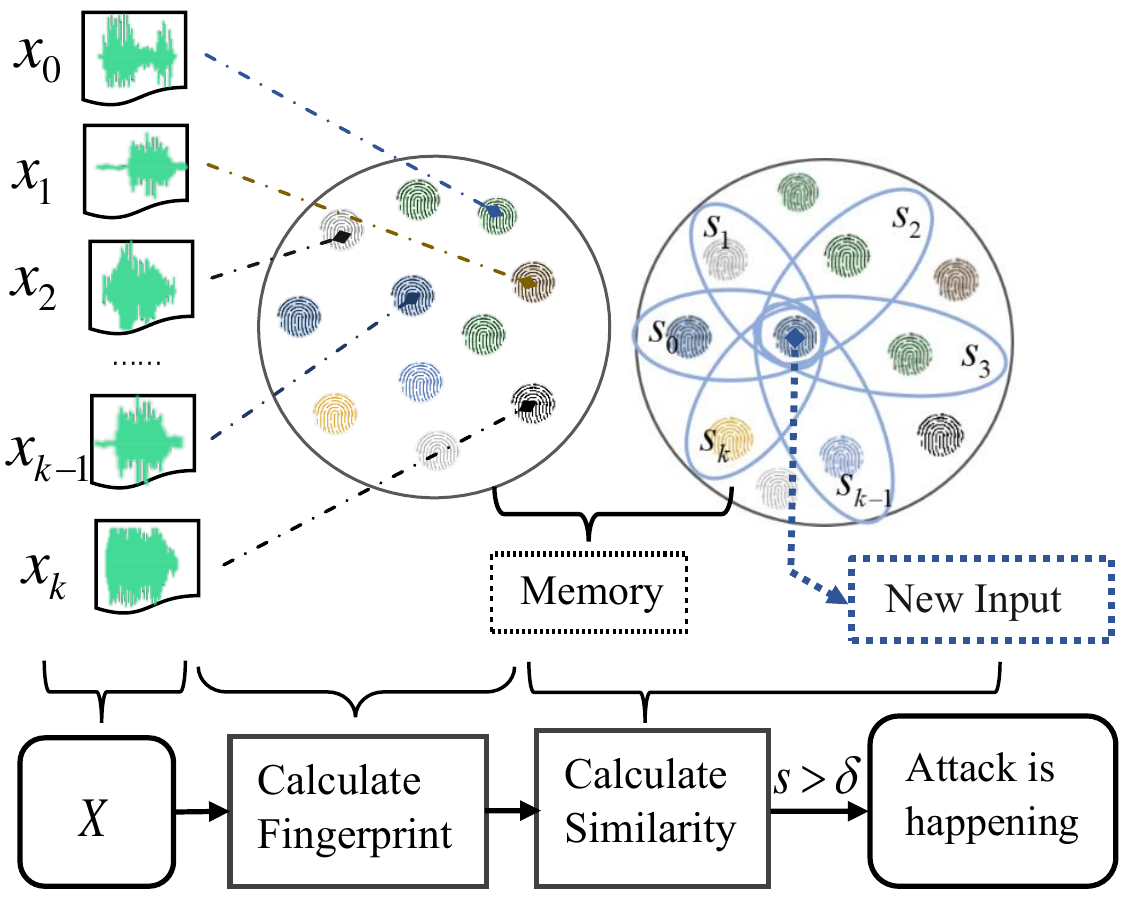}
\caption{\label{frame}  Query-based defense: Architecture for recognizing query-based audio adversarial attack.}
\end{figure}

Our defense architecture is a process-based defense approach and our goal is to find potential attacks in continuous queries. 
Suppose we have determined that the audio fingerprint similarity between the input query and memory is beyond the set threshold, we will report it as part of the attack sequence and take action accordingly.
We can take some actions such as blacklisting the querying user or warning the user.
Fig. \ref{frame} illustrates our scheme.

\;\noindent$\bullet$\textit{Firstly},
place query audio into the cache to form a query memory $\boldsymbol{X}$ of depth $k$. If the number of audio put into the cache is below $k$, consider all queries as a memory sequence. In the process of locating an attack, we expect to consume minimal resources and time, so $k$ should not be too large. Also, it is disadvantageous to discover adversary behavior if  $k$  is too small. The $k$ means the shortest depth before we can make sure that those input queries are intended to produce AEs.

\;\noindent$\bullet$\textit{Secondly},
calculating the fingerprints of all inputs in memory $\boldsymbol{X}$ and overwriting and updating the previous memory.

\;\noindent$\bullet$\textit{Thirdly},
for every new input audio, we calculate the weighted cosine similarity between the new input and each fingerprint in memory. Since audio fingerprint is a particular distribution about time and frequency, the cosine similarity can capture the correlation between such coordinate-dependent distributions. Besides, for each input, there is a necessity to check the legality, so we allocate a weight value  $\alpha$ to each input with the \textit{Inverse Variance Coefficient Method \cite{maths}}. Then, calculate the similarity of the queries via: 
\begin{equation}
\label{s}
\left\{ \begin{array}{l}
s = \sum\limits_{i = 1}^k {{s_i}{\alpha _i}}  \to {s_i}(x,{y_i}) = \frac{{x \times {y_i}}}{{\sqrt {{x^2}} *\sqrt {{y_i}^2} }}\\
\sum\limits_{i = 1}^k {{\alpha _i} = 1} 
\end{array} \right. ,
\end{equation}
where $x$  is the fingerprint of the new input, ${y_i}$ is a fingerprint in memory, and $k$  is the depth of the memory $\boldsymbol{X}$. The final similarity value $s$ is the weighted average value of ${s_i}$.  
The selection of the ${\alpha _i}$ value is explained in the next section. 

\;\noindent$\bullet$\textit{Fourthly},
obtain threshold $\delta $, which implies minimal constraints regarding the input as malicious. When $s > \delta $, it demonstrates that the current input is a potential attempt at generating an AE, and appropriate measures must be taken immediately. 
In practice, for the setting of $\delta$, it is important to have a high uncovering success rate as well as a low false positive ratio. Usually, the false positive ratio will be limited to no more than $10\%$ of the training data, according to the size of the training data set \cite{detecte,26}. The details of $k$ and $\delta$ are explained in the next part of this section.

\textbf{Memory Sequence}

A memory sequence $\boldsymbol{X}$ consists of several queries that are placed in the cache. In the process of attack detection, we expect to consume minimal resources and time. So$\boldsymbol{X}$ should not be too large. Also, it is disadvantageous to detect adversary behavior if  $\boldsymbol{X}$ is too small.  $\boldsymbol{X}$ of depth $k$ means the shortest sequence before we are sure that those queries are intended to produce AEs, and the length of the sequence is $k$, i.e.

\begin{equation}
 \label{SK}
\left\{ \begin{array}{l}
k = \min (f(1),f(2)...f(n))\\
f(i) = \left\{ {\begin{array}{*{20}{l}}
{i,\;if\;f\;{\rm{can}}\;{\rm{detect}}\;{\rm{attacks}}{\rm{.}}}\\
{ + \infty ,\;if\;f\;{\rm{can}}\;{\rm{not}}\;{\rm{detect}}\;{\rm{attacks}}{\rm{.}}}
\end{array}} \right.
\end{array} \right.
 \end{equation}
 
where $f$ is the detection function, ${f(i)}$ indicate whether the function $f$ can detect a sequence of length $i$. Eq. \ref{SK}  implies that the depth of $1,2, ...k-1$ is not sufficient for $\boldsymbol{X}$ to be considered as the intention of generating AEs; depths $k, k+1, ...n$ are considered to be for the purpose of generating AEs,  with the minimum depth is $k$. We explain how to choose the value of $k$ in parameter selection.

\textbf{Query Audio Fingerprints Similarity} \label{Architecture}

The auditory similarity is an important feature in estimating the gap between humans and machines. There is a close auditory similarity between the malicious examples and the benign examples. The malicious examples are produced by appending carefully structured small perturbations to the benign carriers. Although the neural network regards them as two completely different classes, humans believe them as the same intuitively.  So the trait of keeping intuitively consistent with humans is what we need. The audio fingerprint has this trait and is not sensitive as the DNN to perturbations. Fingerprints will maintain high similarity if humans believe they are the same samples.

It is possible to predict whether new input might have a strong correlation with the memory and whether they share the same behavioral attributes, according to the similarity computation between the preserved fingerprints and the new one. This is similar to the recommender system \cite{22,23}, which differentiates users based on their memory behaviors and recommends new content or products \cite{24,25}. 

We note that the digital audio fingerprint \cite{27,28} uniquely flags audio. The small noise of the audio doesn't bother the core information of the fingerprint. And it can defend against some attacks such as audio patching. 
Moreover, it is reliable and feasible in implementation cost to employ fingerprint similarity as an audio similarity. 
Fingerprint similarity relies on the following requirements, assume that $s$ is the similarity function, $x,y,z$ are three candidates in $D$ dimensional space that satisfy Eq. \ref{non}, Eq. \ref{hom}, Eq. \ref{sym}, Eq. \ref{tri}.
\begin{equation}
 \label{non}
s(x,y) \ge 0,(Non - negativity)
 \end{equation}
\begin{equation}
 \label{hom}
s(x,y) = 1,\;only\;x = y.(Homogeneity)
 \end{equation}
 \begin{equation}
 \label{sym}
s(x,y) = s(y,x).(Symmetry)
 \end{equation}
  \begin{equation}
 \label{tri}
s(x,y) + s(x,z) \ge s(y,z).(Triangular{\rm{ }}inequality)
 \end{equation}

A robust acoustic fingerprinting algorithm needs to consider the perception of the audio. When two audio files sound the same, their acoustic fingerprints should be the same or very close, even if there are some differences in their file data.

According to the literature \cite{27, finger}. The fingerprint similarity can be divided into two steps:\textbf{ fingerprint extraction} and \textbf{similarity calculation}.  

\begin{figure}[htb]
\centering
\includegraphics[width=0.48\textwidth]{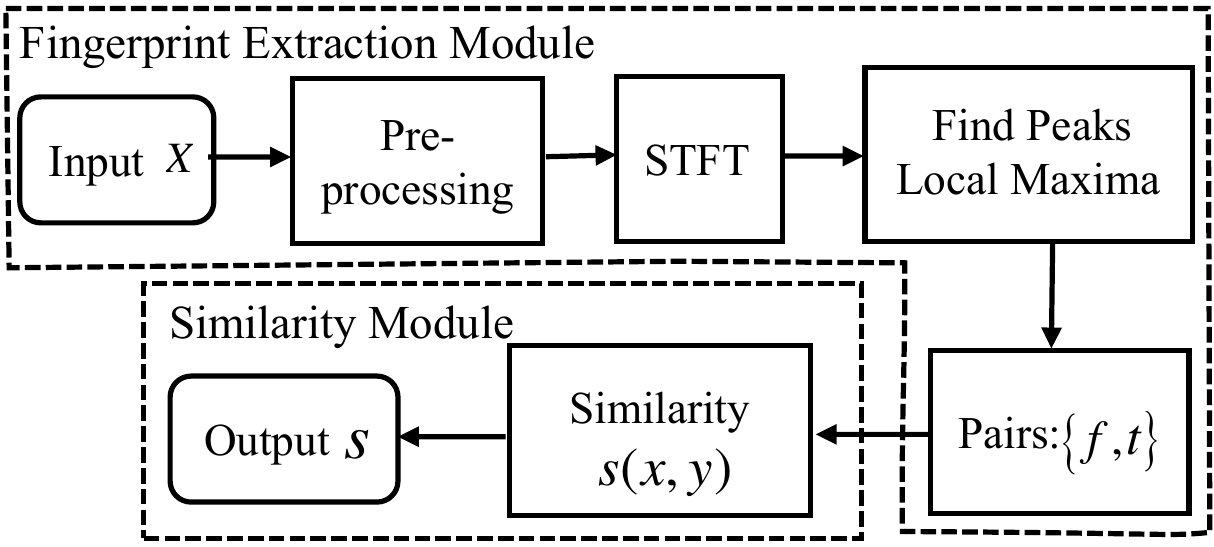}
\caption{ \label{finger}  Architecture of fingerprint similarity calculation.}
\end{figure}

Audio corresponds to a unique fingerprint, so the relationship between digital audio fingerprint $\boldsymbol{F}$ and audio object $\boldsymbol{X}$ is a surjection $h:\boldsymbol{X}\xrightarrow{}\boldsymbol{F}$, and only when $\forall f \in \boldsymbol{F}, \exists x \in \boldsymbol{X},  \to f = h(x)$. That expands to $\left\{ {{x_1} \to {f_1},{x_2} \to {f_2}{\rm{ }}...{\rm{ }}{x_n} \to {f_n}} \right\}$ or  $\{ {f_1} = h({x_1}),{f_2} = h({x_2})...{f_n} = h({x_n})\} $.
For fingerprint ${f_i},{f_j} \in \boldsymbol{F}$, we can obtain similarity ${s_{ij}}$ (${s_{ij}}\in \boldsymbol{S}$)  and $g:\boldsymbol{F}\xrightarrow{}\boldsymbol{S}$ is surjection only when 
$\forall s \in \boldsymbol{S}, \exists {f_i},{f_j} \in \boldsymbol{F}, \to s = g({f_i},{f_j})$. $h,g$ is the map function. 

 \;\noindent$\bullet$\textit{Fingerprint extraction }($h: \boldsymbol{X}\xrightarrow{}\boldsymbol{F}$).
The fingerprint extraction process is illustrated in the fingerprint extraction module in Fig. \ref{finger}. The main procedures include: 

1) Preprocessing: it mainly involves frame split and filtering of the input data.

2) STFT: short-time Fourier transform. For each frame, apply STFT via Eq. \ref{stft}, where ${x(t)}$ is the input signal at time $t$, ${h(t - \tau )}$ is the window function, and $S(\omega ,\tau )$  shows the spectral result if the center of the window function is $\tau$. 

3) Find Peaks: after STFT,  select the frequency peaks $f$ and corresponding time $t$, and make sure the distribution of frequency peaks is uniform.

4) Pairs: pair the obtained frequency peaks $f$ and time $t$, then the result $\left\{ {f,t} \right\}$ is used as fingerprints $f_i$ and $f_i$ is a high-dimensional vector of a certain length.
\begin{equation}
\label{stft}
S(\omega ,\tau ) = \sum\limits_{t =  - \infty }^\infty  {x(t)h(t - \tau ){e^{ - j\omega m}}}   
\end{equation}

 \;\noindent$\bullet$\textit{Find Peaks.}
In Fig. \ref{finger}, after calculating the STFT, we need to uniformly select the peak in the frequency domain. Eq. \ref{cal} describes this process, in which $F(n,m)$ is the two-dimensional matrix after STFT, $H(u,v)$ is the kernel function. Eq. \ref{maxfilter} is the maximum filter and Eq. \ref{thfilter} is the high-pass filter for resetting the frequency to $0$ when the frequency is below the cutoff $D_0$. Both filters are useful for canceling low-frequency components and uniformly capturing the local maximum high frequencies. We choose the former as a tool to find peaks.

 \;\noindent$\bullet$\textit{Similarity calculation }($g:\boldsymbol{F}\xrightarrow{}\boldsymbol{S}$).
After fingerprint extraction, fingerprint $f$ is obtained, which is written as $x = f_i$. Similarly, another fingerprint can be written as $y = f_j$ and its length is the same as $x$.  Then calculate the similarity $s$ between them. The process is illustrated in the Similarity Module in Fig. \ref{finger}. The fingerprint contains coordinate-dependent details. Finally, the similarity of $x$, $y$ could be achieved by Eq. \ref{s}.
\begin{align}
& G(u,v){\rm{ = }}\frac{1}{{NM}}\sum\limits_{n = 0}^{N - 1} {\sum\limits_{m = 0}^{M - 1} {F(n,m)H(u - n,v - m)} }  \label{cal}  \\
& H(u,v) = \mathop {\max }\limits_{s,t \in N(n,m)} [F(s,t)]  \label{maxfilter}\\
& H(u,v) = \left\{ \begin{array}{l}\label{thfilter}
0,{\rm{ }}D(u,v) \le {D_0}\\
1,{\rm{ }}D(u,v) > {D_0}
\end{array} \right.
\end{align}

\textbf{Parameter Selection} \label{para}

\quad\noindent$\bullet$\textit{The choice of $k$ and $\delta$.}\label{k_and_detal}
The larger the $k$ value, the more effective our solution is in observing input queries, and the smaller the $k$ value, the lower the computational cost. The $k$ is the minimum depth of memory before we are sure that those inputs are intended to produce an AE.
The $\delta$ is the minimum similarity before we determine that the current input is malicious.
So the values of $\delta$ would be influenced by the depth of $k$.
Specifically, establishing the threshold requires evaluating fingerprint similarities under the datasets, so that if the entire datasets were to be randomly streamed as queries, $0.1\%$ of the carrier datasets would be marked as attacks. (In theory, the percentage of false positives should be limited to $10\%$ of the dataset size, but since our dataset is small, our value is 100 times smaller than the default.)

\begin{figure}[hthp]
\centering
\centering  
\subfigure[Mini-Librispeech]{
\includegraphics[width=0.227\textwidth]{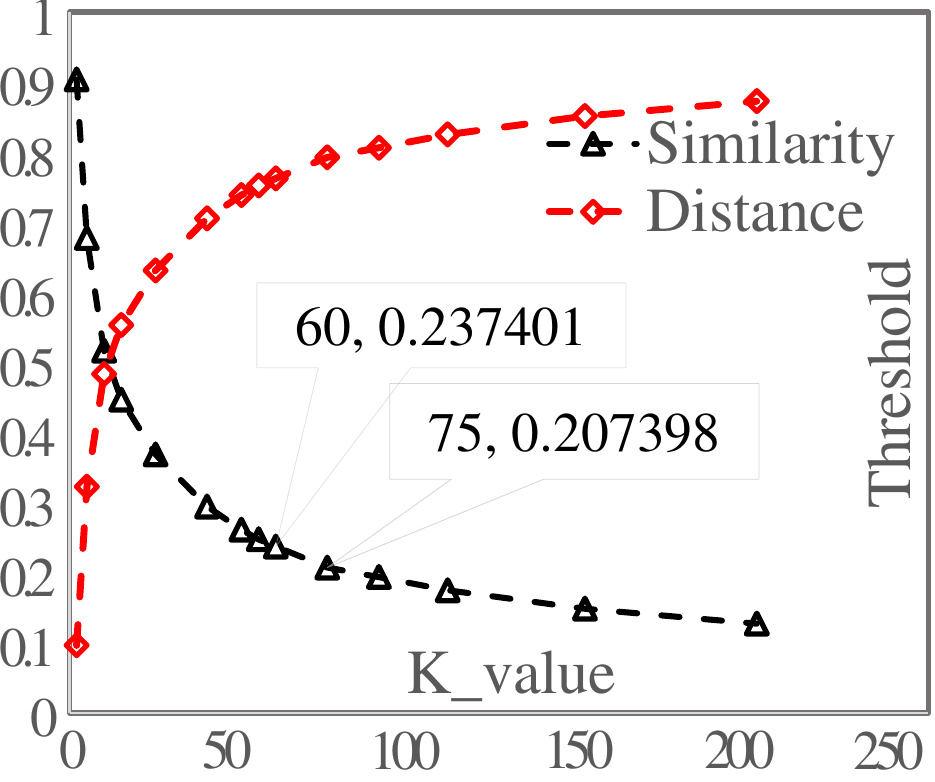}}
\subfigure[Music-sets]{
\includegraphics[width=0.227\textwidth]{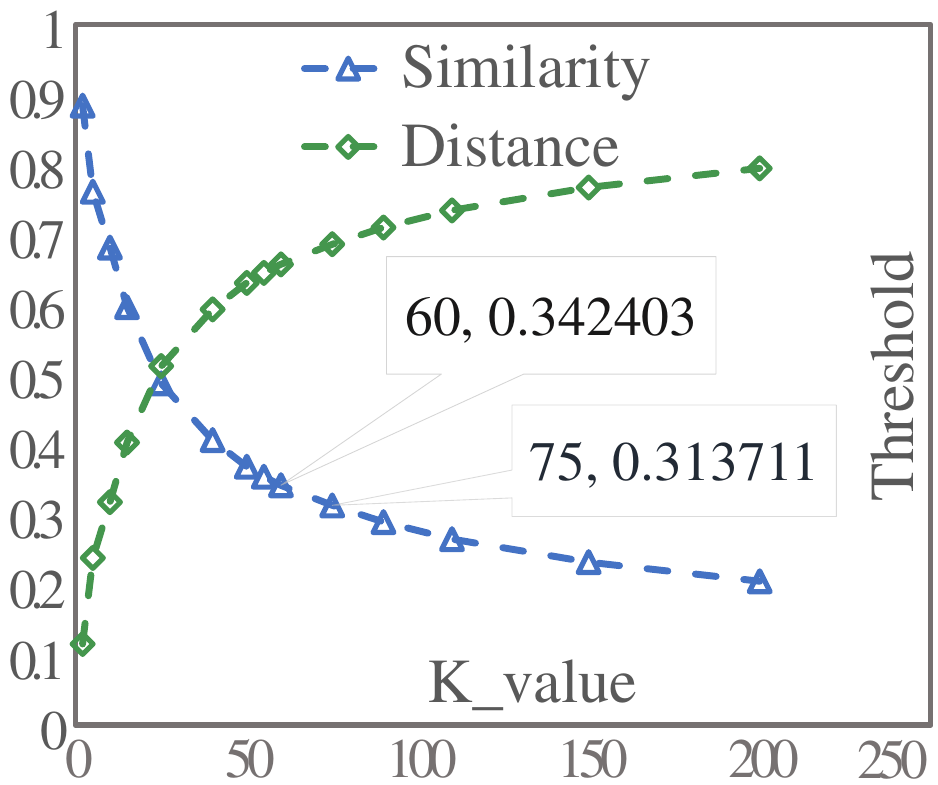}}
\caption{ \label{kdetal}  $k$ and $\delta$: the mean $k$ number of the $0.1\%$ percentile of the datasets as a function of k.}
\end{figure}

Actually, the threshold $\delta$ is a function of $k$, and Fig. \ref{kdetal} discloses their relation. The smaller the threshold $\delta$, the more intense the constraints on the input. Hence small thresholds are advisable, but the too-small value risk regards a benign input as malicious. From what we observed from Fig. \ref{kdetal} with the increase of $k$, the similarity drops sharply in the beginning. (In turn, the distance rises, rapidly. The higher the similarity, the lower the degree of dissociation between input queries, i.e., the closer the distance.) After it reaches around $k=75$, curves become smooth and increase modestly with $k$, and the process is quite gentle, so we set up $k$ as $75$ and the thresholds $\delta$ in both datasets are $0.313711$ and $0.207398$.

 \;\noindent$\bullet$\textit{The choice of $\alpha$.}\quad
\label{fake}
First, let's consider a case, in Eq. \ref{eq_fake} below:
\begin{equation}
\label{eq_fake}
\left\{ \begin{array}{l}
{{\rm{s}}_1}{\rm{  = f(}}{{\rm{X}}_A}{\rm{, }}{{\rm{q}}_{new}}{\rm{), }}{{\rm{X}}_A} = \{ {{\rm{q}}_0}{\rm{,}}{{\rm{q}}_1}...{{\rm{q}}_m}{\rm{,}}{{\rm{q}}_n}\} \\
{{\rm{s}}_2}{\rm{  = f(}}{{\rm{X}}_B}{\rm{,}}{{\rm{q}}_{new}}{\rm{) , }}{{\rm{X}}_B} = \{ {{\rm{q}}_0}{\rm{,}}{{\rm{q}}_1}...{\rm{p,}}{{\rm{q}}_n}\} 
\end{array} \right.
\end{equation}
There exist two memory sequences $\boldsymbol{X}$, where memory $X_A$ consists of $\{ {{\rm{q}}_0}{\rm{,}}{{\rm{q}}_1}...{{\rm{q}}_m}{\rm{,}}{{\rm{q}}_n}\} $ and $X
_B$ is: $\{ {{\rm{q}}_0}{\rm{,}}{{\rm{q}}_1}...{\rm{p,}}{{\rm{q}}_n}\} $, ${{\rm{s}}_1}$ and ${{\rm{s}}_2}$ are the similarity of the two sequences with new input, $f$ is the fingerprint similarity function. The key distinguishing element between $X_A$ and $X_B$ is that the query ${q_m}$ differs from $p$. Assuming that $p$ is a query deliberately placed in the queries by an adversary. The adversary's purpose of injecting $p$ is to try to fabricate a fake input (i.e., almost irrelevant to the former) to confuse the analysis of the similarity and hide her intent. Essentially, both $X_A$ and $X_B$  are malicious memory sequences with only trivial disparity. But ${{\rm{s}}_1}$ is below the threshold and ${{\rm{s}}_2}$ is beyond the threshold, $X_A$ is decided as a potential attack while $X_B$ is not decided as a potential attack due to the injection of a fake query. We call this $p$-input as ``\textit{fake query}'', and the ratio of ``\textit{fake query}'' to all queries is called ${p_{fake}} $ (${p_{fake}}{\rm{ = (p/k)*100\% }}$). In our experiments, we found that the $s$ value would change sharply when there were ``fake queries'' in the query memory and we employed the \textit{Inverse Variance Coefficient Method} \cite{maths} to describe such fluctuations and disparities. According to this method, it is easy to determine the weights $\alpha $, which are assigned as follows: 
\begin{equation}
{\alpha _i} = 1 - \frac{{st{d_i}(s(l))}}{{men{a_i}(s(l))}} \to {\alpha _i} = \frac{1}{\alpha }{\alpha _i}(\alpha {\rm{ = }}\sum\limits_{i = 0}^k {{\alpha _i})} ,
\end{equation}
where $men{a_i}$ depict the mean, $st{d_i}$ depict the standard deviation, and $s(l){\rm{ }}$ depict getting the query vector of length $l/2$ before and after the i-query.
For $l$, we set the maximum value as $7$ (No more than $10\%$ of the memory length, i.e. ${l_{\max }} = floor(0.1*k) = floor(0.1*75) = 7$) and  $l$ begin with 2 (The mean and variance are worthwhile at least two values). Then, the value increases linearly. When it exceeds the maximum value, $l$ shrinks to half of the original value and then increases linearly duplicate. Repeat this process until all elements are traversed.

\begin{table*}[tbp]\small
\centering
 \caption{\label{tab:1}Non-adaptive attack evaluation. SRoA denotes the success rate of attack. The higher the value of $DSR$ and $FSNR$, the more beneficial. Normally, every $k$ ($k$=75) query is detected once, and if the queries are less than $k$, at least one detection is performed for all $n$ queries, and the ratio of $n/k$ is the detections.}
 {
\begin{tabular}{c c c c c c c}
\toprule[1.5pt]
\textbf{Attack} &\textbf{dataset} & \textbf{SRoA($\%$)}  &\textbf{ Avg.Queries($n$)} &\textbf{ Detections}&\textbf{$\boldsymbol{DSR}$($\%$) } &\textbf{$\boldsymbol{FSNR}$ (dB)}\\ 
\toprule[1.2pt]
\textbf{CS} & Music-sets &\makecell[c]{ 100.00} & \makecell[c]{$\sim$300} & \makecell[c]{$\sim$3.92} & \makecell[c]{98.00}& \makecell[c]{7.38}  \\
\textbf{DW} & Music-sets & \makecell[c]{98.00} & \makecell[c]{$\sim $150}& \makecell[c]{$\sim $1.7} & \makecell[c]{84.74} & \makecell[c]{18.41}  \\  
\toprule[0.5pt]
\multicolumn{2}{c}{\textbf{Average}} & \makecell[c]{100.00} & \makecell[c]{$\sim $225}& \makecell[c]{$\sim$2.81} & \makecell[c]{91.37} & \makecell[c]{12.90}  \\  
\toprule[0.5pt]
\textbf{IRTA}  & Mini-librispeech & \makecell[c]{100.00} & \makecell[c]{$\sim $5000} & \makecell[c]{$\sim $56.00} & \makecell[c]{84.00} & \makecell[c]{40.97} \\
\textbf{DS } & Mini-librispeech & \makecell[c]{100.00} &\makecell[c]{$\sim $1000} & \makecell[c]{$\sim $11.00} & \makecell[c]{82.50} & \makecell[c]{13.02} \\
\toprule[0.5pt]
\multicolumn{2}{c}{\textbf{Average}} & \makecell[c]{100.00} & \makecell[c]{$\sim $3000}& \makecell[c]{$\sim$34.00} & \makecell[c]{83.25} & \makecell[c]{27.00}  \\  
\bottomrule[1.5pt]
\end{tabular}
}
\end{table*}

\section*{Evaluation}\label{eval}

In this section, we will show the evaluation results of our scheme for some non-adaptive attacks and adaptive attacks. We collected open-source code attacks as much as possible, and we did not evaluate attacks without open-source code, but we made some surveys about their details. Finally, we evaluated four class attacks that are well-known in the audio adversarial attack. Those are sufficiently representative and the bulk of the other work revolves around them. We evaluate the CommanderSong (CS) \cite{7} attacks and the Devil's Whisper (DW) \cite{9} attacks by applying the Music-set. The Mini-Librispeech dataset is applied to assess the IRTA\footnote {IRTA  is an abbreviation for the attack of the paper ``Imperceptible, Robust, and Targeted Adversarial Examples for Automatic Speech Recognition''} attack \cite{5} and DS\footnote {DS is an abbreviation for the attack of  ``Audio Adversarial Examples: Targeted Attacks on Speech-to-Text''. This paper attacks the DeepSpeech model.} attack \cite{4}. Those attacks all reported a success rate of attacks (SRoA) of almost $100\%$.

\textbf{Datasets}

Our scheme conducts experiments on Mini-Librispeech \footnote{\url{https://www.openslr.org/31/}} and Music-sets  datasets (We build a carrier library of music-based samples containing 10,553 music clips. Appendix \ref{music-sets} contains all details about Music-sets).
For Mini-Librispeech, this is a dialog-based dataset that some classic attack works rely on it and we cannot ignore it \cite{6, API65,multi-objective}. 
For Music-sets, music has the characteristic of large-scale availability in  most situations, and its accessibility and popularity allow it to become a candidate of the carrier in attacks. Lots of strong attacks refer to music as the necessary carrier for producing AEs. \cite{7,4,9,34,15} 
So, defense and evaluation of the AEs on musical carriers are inevitable and important. 

\textbf{Evaluation Metric} \label{metric}

\quad\noindent$\bullet$\textit{$DSR$.}
To evaluate the effectiveness of our approach for defending the query-based attacks, we employ the detection success rate ($DSR$) and First-Signal-to-Noise Ratio ($FSNR$) as the evaluation metrics. The detection success rate ($DSR$) is the most intuitive metric to evaluate the detection results. To calculate it as follows:
\begin{equation}
DSR({\% }) = \frac{{{\operatorname{d} _n} * k}}{{{a_n}}} * 100\%  ,
\end{equation}
where ${\operatorname{d} _n}$ is the number of  detections, ${a_n}$ is the number of queries, and $k$ is the length of memory $X$. Obviously, the $DSR$ value is below 1 because ${a_n} > {{{d} _n} * k}$ is clear. The detection occurs after performing at least one query. For our purposes, we consider it to measure the probability of finding adversary behavior. A higher $DSR$ is preferable.  

 \;\noindent$\bullet$\textit{$FSNR$ .}
The First-Signal-to-Noise Ratio ($FSNR$) is a function that defines the minimum $SNR $ to detect an attack, i.e., how much $SNR$ when we can detect the attack, as shown in Eq. \ref{FSNR}: 
\begin{equation}
\label{FSNR}
FSNR(dB) = 20{\log _{10}}(\frac{{{A_x}}}{{F{A_\delta }}}), 
\end{equation}
where $x$ is the original sound, $\delta$ is the perturbation, ${A_x}$ is the amplitude of the original sound, and $F{A_\delta }$ is the amplitude of the perturbation when the first attack is detected. This is a metric of the relative value of distortion of the AE vs the original sound. The higher $FSNR$ describes that the query will be regarded as a suspect under a smaller perturbation.

\textbf{Non-adaptive Attack Evaluation}

We evaluate four class attacks that are well-known in the audio attack. Those are sufficiently representative and the bulk of the other work revolves around them. We evaluate the CommanderSong (CS) \cite{7} attack and the Devil's Whisper (DW) \cite{9} attacks by applying the Music-set. The Mini-Librispeech dataset is applied to assess the IRTA attack \cite{5} and DS attack \cite{4}. Those attacks all reported a success rate of attacks (SRoA) of almost $100\%$.
CS attack is the representation of employing music as carriers and some subsequent work\cite{9,15} set it as an indispensable collection.
The DW attack is the typical instance for commercial black-box APIs. Subsequently, much of the work\cite{API65,API66} on black-box attacks has to test on APIs.
IRTA attack based on the psychoacoustic hiding model is an outstanding work of the period. And several studies\cite{34,59} adopted the psychological masking effect.
DS attack is the earliest version of voice attack, which launched the gateway to voice attack and provided a reliable infrastructure for the subsequent works.

\quad\noindent$\bullet$\textit{N1. CS attack Evaluation}.
CS attack is a white-box attack by injecting target commands into the song. It started a precedent of producing AEs with music as a carrier and achieving a $100\%$ success rate of attacks (SRoA) on the Kaldi speech recognition system. It has a profound influence, and many follow-up works set it as an indispensable reference. For the defense based on our approach, there are few blanks in the music, the spectrum is abundant, and the fingerprints are often more reliable than those of the dialogue version. Tab. \ref{tab:1} shows that CS examples spend an average of about 300 visits to the target model. Our security architecture can accurately detect such attacks with $DSR$ up to $98\%$. However, the value of $FSNR$ is only 7.38 dB, revealing that the AEs were already very noisy when we suspected the query was an attack. The primary factors of this situation are that the small perturbation is not ideal for a CS attack and the perturbation is constrained to a very broad range. Therefore, the amount of additional noise is significant.
Apart from that, various audio lengths will affect the SRoA of AE. To ensure the validity of AE, the length of audio ought to be no shorter than 4s. The longer the audio, the richer the fingerprint, which is more helpful for detection. However, the shorter audio is not beneficial for the adversary to generate AEs successfully.

 \;\noindent$\bullet$\textit{N2. DW attack Evaluation.}
DW attack first accomplished a black-box attack on commercial speech recognition APIs (including Google Assistant, Google Home, Amazon Echo, and Microsoft Corina). Since then, attacks on APIs have gradually become a necessary option for black-box attacks and the most intuitive indicator of the attack algorithm. Tab. \ref{tab:1} shows that DW also works based on the music dataset, which accounts for $50\%$ of CS in the average query to the target model and  SRoA is close to $98\%$. On defense, our approach enables a $DSR$ of $84.74\%$ under DW attack. DW attack employs a local substitution model to simulate approximately the target model of the APIs ASR system. It helps to diminish the number of queries and the likelihood of triggering detection. So $DSR$ possible losses. The $FSNR$ value is $18.41dB$, which is about 2.5 times that of CS. DW increases the $FSNR$ value by reducing the number of visits to the model, and the perturbation naturally decreases.

DW adopts Noise Model to augment the physical robustness of AEs. However, the SRoA is deeply relevant to the environment and the device. Regarding the noise model, the combination of our scheme with some straightforward measures (e.g., down-sampling, filtering) can raise the level of difficulty of physical attack.

\begin{table*}[thb]
{
\caption{\label{tab-ASR-attack} An overview of the query-based attacks against ASR. \textbf{ Note:} in the table, ``GD'', ``GA'', ``GE'', ``SGE'' represent the Gradient Descent, Genetic Algorithm,  Gradient Estimation, and Selective Gradient Estimation. ``Alt-M'', ``Psy-M'', ``Co-E'', ``PSO'', ``Mul-Obj GO'' represent the Alternative Models, Psychoacoustic Masking, Co-evolutionary algorithm, Particle Swarm Optimization, Multi-Objective Genetic Optimization. ``M or D''  represents the Music-carrier or Dialogue-carrier, ``-'' denotes the author didn't show, and  ``*'' denotes the author told us the WER of the attack model to AEs was increased to $980\%$.}
\resizebox{1.\textwidth}{!}{%
\begin{tabular}{cccccccc}
\toprule[1.5pt]
\textbf{Attack} & \textbf{Task} & \textbf{Attack Method} & \textbf{Attack Model} & \textbf{Target} & \textbf{M or D} & \textbf{Avg.Queries} & \textbf{SRoA(\%)} \\
\toprule[1.2pt]
\textbf{CS\cite{7}} & ASR & GD & Kaldi-Aspire & \begin{tabular}[c]{@{}c@{}}Play music.\\ Open the front door.\\ Turn off the light.\end{tabular} & M & $\sim$300 & 100 \\
\textbf{DS\cite{4}} & ASR & GD & DeepSpeech & \begin{tabular}[c]{@{}c@{}}Okay google browse \\ to evil dot com.\end{tabular} & M \& D & $\sim$1000 & 100 \\
\toprule[0.5pt]
\textbf{DW\cite{9}} & ASR & Alt-M & APIs & \begin{tabular}[c]{@{}c@{}}Turn off The Light\\ Take a picture.\\ Call 911.\end{tabular} & M & $\sim$150 & 100 \\
\toprule[0.5pt]
\textbf{DSG\cite{6}} & ASR & GA \& GE & DeepSpeech & \begin{tabular}[c]{@{}c@{}}Morning body.\\ Ball charge.\\ More they.\end{tabular} & D &  $\sim$150000 & 35 \\
\textbf{Foolgle\cite{API65}}& ASR & GA & Google-API & - & D & - & 86 \\
\toprule[0.5pt]
\textbf{SGEA\cite{reduce-query}} & ASR & SGE & DeepSpeech & \begin{tabular}[c]{@{}c@{}} Thank you.\\ Hello world.\\ Open the door.\end{tabular} & D & $\sim$78000 & 98 \\
\toprule[0.5pt]
\textbf{IRTA\cite{5}} & ASR & Psy-M & Lingvo & \begin{tabular}[c]{@{}c@{}}Old will is a fine fellow \\ but poor and helpless sin\\ -ce missus rogers had her\\ accident.\end{tabular} & D & $\sim$5000 & 100 \\
\textbf{PHA\cite{34}} & ASR & Psy-M & Kaldi-WSJ & \begin{tabular}[c]{@{}c@{}}Do not blame you.\\ The command is planted.\\ The cake is a lie.\end{tabular} & M \& D & $\sim$500 & 98 \\
\textbf{EPA\cite{59}} & ASR & Psy-M & \begin{tabular}[c]{@{}c@{}}DeepSpeech \\ and Wav2Letter\end{tabular} & \begin{tabular}[c]{@{}c@{}}That is comparatively nothing.\\ Talking later is beneath us.\\ But there seemed no.\end{tabular} & D & $\sim$1000 &  76\\
\toprule[0.5pt]
\textbf{Occam\cite{15}} & ASR & Co-E & \begin{tabular}[c]{@{}c@{}}DeepSpeech \\ and APIs\end{tabular} & \begin{tabular}[c]{@{}c@{}}Call my wife.\\ Navigate to my home.\\ Open the door.\end{tabular} & M \& D & $\sim$30000 & 100 \\
\toprule[0.5pt]
\textbf{SirenAttack\cite{gen}} & ASR & PSO & DeepSpeech & \begin{tabular}[c]{@{}c@{}}Read last sms from boss. \\ Call the police for help.\end{tabular} & D & $\sim$1000 & 100 \\
\toprule[0.5pt]
\textbf{MOGA-Attack\cite{multi-objective}} & ASR & \begin{tabular}[c]{@{}c@{}}Mul-Obj\\ GO\end{tabular} & \begin{tabular}[c]{@{}c@{}}DeepSpeech \\ and Kaldi\end{tabular} & \begin{tabular}[c]{@{}c@{}}A cat.\\ All of these.\\ That i love you.\end{tabular} & D & - & *\\
\bottomrule[1.5pt]
\end{tabular}
}
}
\end{table*}

 \;\noindent$\bullet$\textit{N3. IRTA attack Evaluation.}
IRTA attack is a two-stage attack algorithm on Lingvo, concealing target commands to a space that the human ear cannot hear through a psychoacoustic masking model.
The IRTA example is based on the open-source dataset Librispeech. This type of dialogue audio contains a large number of silent fragments. Therefore, the fingerprint of the audio is inferior to that of the music. But the inspiring thing is that our approach maintains a robust attack detection and that the $DSR$ reaches $84\%$. This can be attributed to the time cost of this type of attack (Producing a successful adversarial example costs 24.8h) leads to a remarkable number of queries. Such massive queries easily provoke the inspection of the defense system. Moreover, the perturbation is very small, and the $FSNR$  can reach $40.97$dB  in which the psychoacoustic masking model plays an important role. Still, the perturbation would reflect the frequency domain and the fingerprint extraction happens in the frequency domain. We can further presume that it will be costly to bypass our defenses for adversaries with an emphasis on hidden perturbation via psychoacoustic masking.
Nevertheless, it also exposes a critical concern: \textit{In the areas that humans fail to hear, is there a necessity for the machine to do so?} AI researchers aim to narrow the gap between humans and machines, so machines should also appear human-like for regions beyond human perception. Blocking such attacks implies that the machine does not have the power to do anything in the regions where humans are unable to perceive, thus, the attack will completely dissolve.

 \;\noindent$\bullet$\textit{N4. DS attack Evaluation.}
DS attack is a type of attack first implemented on DeepSpeech. At its core is to optimize the CTC-Loss function. Compared to IRTA attacks, DS is relatively heavily perturbed that maybe without applying the theory of psychological masking, and relatively poorer $FSNR$ but $DSR$ is $82.5\%$ closer to IRTA. Compared to CS and DW attacks, DS and IRTA attack are implemented on Librispeech containing rare fingerprint information, so $DSR$ is inferior to CS and DW. Nevertheless, the general $FSNR$ is superior to the former, showing the method's detection capability to attacks with small perturbations.
Separate work deploys genetic algorithms and gradient estimation to generate adversarial samples. However, gradient estimation relies on the sampling theory. Biological evolutionary algorithms demand substantial expenses without the guideline of the gradient. 
The literature \cite{6}  queries numbers up to 1000+, and the literature \cite{15}  reach a stunning 30000+. 
From Tab. \ref{tab:1}, it has a remarkably higher detection rate for query numbers above 1000+. 
Multiple query numbers are an obvious disadvantage of the evolutionary algorithm. Unless improving this shortcoming, do not expect to evade our inspection.

We investigated the perturbation level of AEs so that we can easily compare them with $FSNR$, as shown in Tab. \ref{tab:3}.    

\begin{table*}[htbp]\small
\centering
\captionsetup{justification=centering}
\caption{\label{tab:2}$DSR$ as a function of the ${p_{fake}}$. The effect of different false query ratios on the success rate of detection.}
 {
\begin{tabular}{c c c c c c c}
\toprule[1.5pt]
\textbf{Attack/(Sets)} & \makecell[c]{\textbf{$\boldsymbol{DSR}$(\%)}\\\textbf{$\large{p_{fake}}$=0\%}} & \makecell[c]{ \textbf{$\boldsymbol{DSR}$(\%)}\\$\large{p_{fake}}$\textbf{=10\%}} & \makecell[c]{\textbf{$\boldsymbol{DSR}$(\%)}\\$\large{p_{fake}}$=\textbf{25\%}} & \makecell[c]{\textbf{$\boldsymbol{DSR}$(\%)}\\$\large{p_{fake}}$=\textbf{40\%}} & \makecell[c]{\textbf{$\boldsymbol{DSR}$(\%)}\\${p_{fake}}$=\textbf{50\%}} & \makecell[c]{ \textbf{$\boldsymbol{DSR}$(\%)}\\$\large p_{fake}$=\textbf{60\%}}\\
\toprule[1.2pt]
\textbf{CS (Music-sets)}&\makecell[c]{ 98.00 }&\makecell[c]{ 98.00 }&\makecell[c]{98.00 }& \makecell[c]{76.20 }& \makecell[c]{76.20} &\makecell[c]{5.44 }\\
\textbf{DW (Music-sets) } &\makecell[c]{ 84.74 }& \makecell[c]{79.66} & \makecell[c]{79.66} & \makecell[c]{57.20} & \makecell[c]{33.47} &\makecell[c]{ 0.00} \\
\toprule[0.5pt]
\textbf{Average}&\makecell[c]{ 91.36 }&\makecell[c]{ 88.82 }&\makecell[c]{88.82}& \makecell[c]{66.70 }& \makecell[c]{54.84} &\makecell[c]{2.72 }\\
\toprule[0.5pt]
\textbf{IRTA (Mini-Librispeech) } &\makecell[c]{ 84.00 }&\makecell[c]{ 84.00} &\makecell[c]{84.00} &\makecell[c]{84.00}&\makecell[c]{ 84.00 }&\makecell[c]{ 3.00} \\
\textbf{DS (Mini-Librispeech)}   &\makecell[c]{82.50 }& \makecell[c]{81.75} &\makecell[c]{ 81.75} & \makecell[c]{81.00} & \makecell[c]{80.25}&\makecell[c]{0.00}\\
\toprule[0.5pt]
\textbf{Average}&\makecell[c]{ 83.25 }&\makecell[c]{ 82.88 }&\makecell[c]{82.88}& \makecell[c]{82.50 }& \makecell[c]{82.13} &\makecell[c]{1.50 }\\
\bottomrule[1.5pt]
\end{tabular}
}
\end{table*}

\begin{table}[htp]\small
\label{constrain}
 \centering
\caption{\label{tab:3} Perturbation levels for different attacks (The numbers in the table are the outcome after normalization). Shows perturbation levels for different attacks. The higher the level of perturbation, the smaller the $FSNR$.}
 {
 \resizebox{0.45\textwidth}{!}{
\begin{tabular}{cccc}
\toprule[1.pt]
\textbf{Constraint}  & \textbf{$|{\rm{|}}\delta |{|_1}$} & \textbf{$||\delta |{|_2}$} & \textbf{$||\delta |{|_\infty }$} \\
\toprule[1.pt]
\textbf{CS-Attack} & 346.85 & 23.62 & 0.24 \\
\textbf{DW-Attack} & 198.21 & 2.60 & 0.05 \\ 
\toprule[0.5pt]
\textbf{Average} & 272.53 & 13.11 & 0.15 \\  
\toprule[0.5pt]
\textbf{IRTA-Attack} & 169.58 & 0.80 & 0.02 \\
\textbf{DS-Attack} & 63.09 & 0.37 & 0.37 \\
\toprule[0.5pt]
\textbf{Average} & 116.34 & 0.59 & 0.20\\
\bottomrule[1.pt]
\end{tabular}
}}
\end{table}

 \;\noindent$\bullet$\textit{N5. Other query-based attacks evaluation.}
Other query-based attacks, the majority of them are based on the 4 attacks above. CS attack is the representation of employing music as the carrier. After that, subsequent work \cite{9,15} also set it as an indispensable collection. The DW attack is a typical example of attacking commercial black-box APIs. Subsequently, a lot of the work \cite{API65,API66} on black-box attacks has to be tested on APIs. IRTA attack based on the psychoacoustic hiding model is an outstanding work of the period. Several studies \cite{34,59} adopted the psychological masking effect. Literature \cite{reduce-query, gen} using biological evolutionary algorithms to perform attacks and optimize the number of queries. DS attack is the earliest relatively sophisticated version of an audio attack, which provides a reliable infrastructure for subsequent works.
Since our defense framework is process-based, we were unable to evaluate the attacks without open-source code but still surveyed them. More relevant details are provided in Tab. \ref{tab-ASR-attack}.

We can learn from the above that applying a music carrier is quite advantageous for detection, also the detection is significant when the number of queries is numerous. The critical factor is that the fingerprints of music are more obtuse to perturbations, while the conversational ones are not. 
In terms of fingerprint extraction, Fig. \ref{c-ae-f} from the Appendix \ref{fingerprint} supports similar results. 
In the following, we built a more robust defense system that raises the average DSR beyond $90\%$ and substantially strengthens our defense, Tab. \ref{tab-noise-aug} shows the results.
For adversaries, unless improving those shortcomings, do not expect to evade our inspection.
Below, we propose a more robust defense system by combining other methods, which can achieve a detection ratio of over $90\%$, The details are in \ref{roboust}.

\begin{table*}[htb]

\centering
\caption{\label{tab-noise-aug} Robust defense: we add noise based on different $SNR$, the lower the SNR, the heavier the added noise. }
 {
\resizebox{0.85\textwidth}{!}{%
\begin{tabular}{cccccc}
 \toprule[1.pt]
\textbf{SNR(dB)} & \textbf{CS-Attack} & \textbf{DW-Attack} & \textbf{IRTA-Attack} & \textbf{DS-Attack} & \textbf{Average} \\ \hline
\toprule[0.8pt]
150 & 3.92/98.00 & 1.70/84.74 & 56.00/84.00 & 11.00/82.5 & 17.44/87.31 \\
100 & 3.92/98.00 & 1.70/84.74 & 56.50/84.75 & 11.00/82.50 & 17.57/87.50 \\
\textbf{75} & \textbf{3.92/98.00} & \textbf{1.70/84.74} & \textbf{58.50/87.75} & \textbf{12.50/93.75} & \textbf{19.16/91.06} \\
\textbf{50} & \textbf{3.92/98.00} & \textbf{1.70/84.74} & \textbf{63.00/94.50} & \textbf{13.00/97.50} & \textbf{20.41/93.69} \\
\textbf{25} & \textbf{3.92/98.00} & \textbf{1.70/84.74} & \textbf{63.00/94.50} & \textbf{13.00/97.50} & \textbf{20.41/93.69} \\
0 & 3.70/92.50 & 1.50/75.00 & 10.00/15.00 & 3.00/22.50 & 4.55/51.25 \\ 
\bottomrule[1.pt]
\end{tabular}
}
\begin{tablenotes}
\footnotesize
 \item[]  \textbf{Note}: ``3.92/98.00'' indicates that the average queries are 3.92, and $DSR$ is $98.00\%$
\end{tablenotes} 
}
\end{table*}

\textbf{Adaptive Attack Evaluation}

Whereas our defense framework can effectively detect existing attacks, it only assures in ``\textit{zero-knowledge}'' attack scenarios where the attacker is unknown of the existence of the defense framework. In order to reliably implement our framework in practice, we have to assess adaptive adversaries who understand the defense details entirely and intend to deploy some strategies to bypass the defense mechanism. Following the guidelines of \cite{30}, we designed adaptive attacks to evaluate the ability of our defense to adaptive attacks. 
According to the defense details we consider both adaptive attacks: \textbf{\textit{Random Noise attack}} and\textbf{ \textit{Proportion of Fake Queries attack}}.

\begin{figure}[tpb]
\centering  
\includegraphics[width=0.45\textwidth]{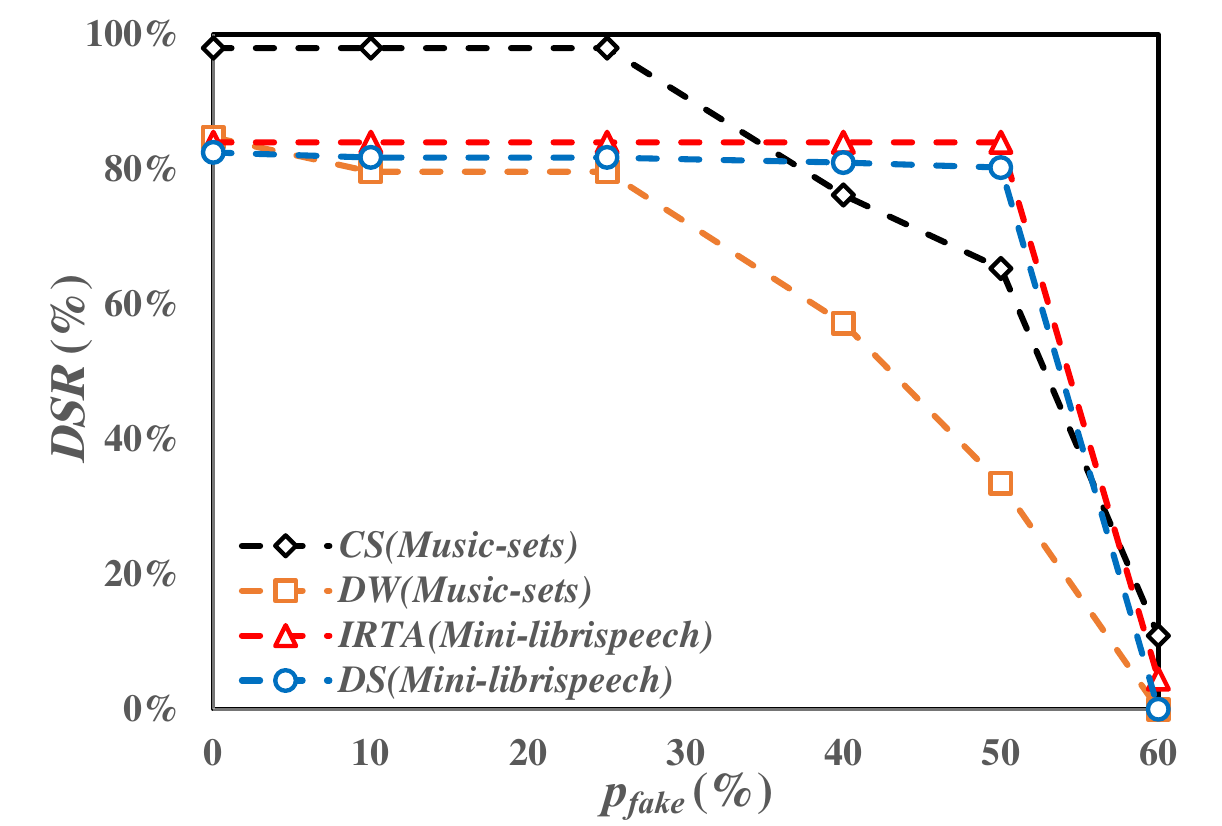}
\caption{\label{fig-fake} $DSR$ as a function of the ${p_{fake}}$.}
\end{figure}

\begin{figure}[hbt]
 \centering
\subfigure[Noise-snr impact  $DSR$.]{
\includegraphics[width=0.225\textwidth]{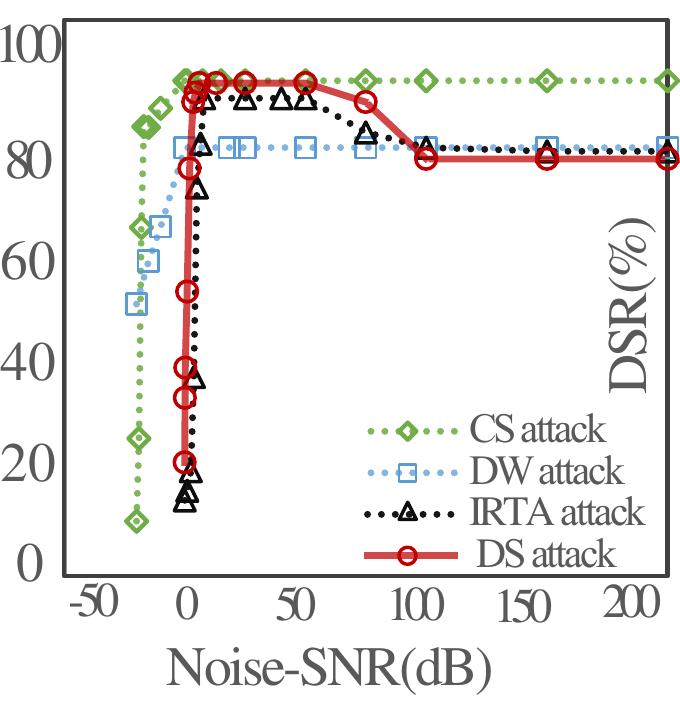}}
\subfigure[Noise-snr impact $SRoA$.]{
\includegraphics[width=0.225\textwidth]{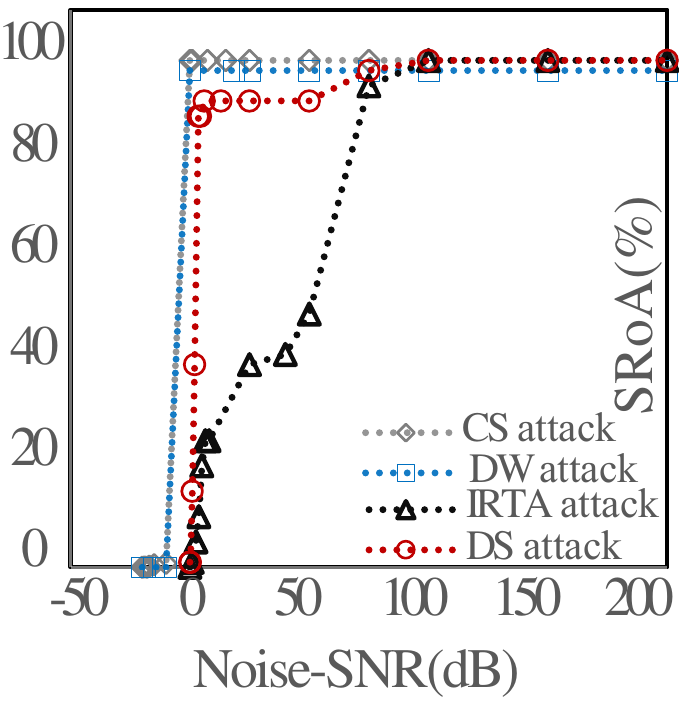}}
\caption{\label{noise}  Adaptive attack: Different noise-snr to disturb the extraction of fingerprints. Noise-SNR indicates the noise of different SNR. The smaller Noise-SNR means higher noise level.}
\end{figure}

Fig. \ref{fig-fake} shows the effect of ${p_{fake}} $ on $DSR$.

 \;\noindent$\bullet$\textit{\textit{A1. Random noise attack.}}
 We conceive an adaptive attack of corrupting fingerprint extraction. Randomly insert noise with different SNR to the audio in the process of query. Forcing the ${x^ * }$ to bypass the defense, and successfully attack the ASR, and the perturbation is not easily perceived by the human.   
In Fig. \ref{noise}, according to audio quality theory, when $SNR$ is above $70$, it belongs to high-fidelity quality audio. When $SNR=0$, the noise has the same energetic value as the original audio, so when  $SNR$ is below $ 0$, the original audio is almost flooded with noise.  

As shown in Fig. \ref{noise}, for CS and DW attacks, when Noise-SNR is below 0, the SRoA is also nearly 0. Therefore the malicious queries are almost unsuccessful in attacking the ASR system, which is unacceptable for the adversaries. When Noise-SNR$>$0, the SRoA and $DSR$ are rapidly recovering to their maximum value and keep it and the SRoA, in other words, $DSR$ displays a comparable consistency. Though large noise decreases the $DSR$ value but also decreases SRoA, which diverts from the adversary's target. So it is impossible to achieve superior SRoA while trying to break our defense. 
However, when the Noise-SNR value gradually increases, for IRTA and DS attacks, SRoA is rapidly recovering to its maximum value and keeping it except IRTA attack recovery is slower and the $DSR$ value sharply rises and then gently drops until it becomes peaceful. Since Mini-Librispeech is a dialogue-based dataset and it contains a lot of blank frames, when inserting noise, it will fill the blank and become more helpful to the extraction of fingerprints. It can be deduced that joining appropriate noise can improve the robustness to our method. The query of containing noise does not undermine our defenses, on the contrary, it leads the defense system more sensitive and robust.

 \;\noindent$\bullet$\textit{\textit{A2. Proportion of Fake Queries Attack.}}
Moreover, we noted above that some adversaries use ``\textit{fake queries}'' to develop a fake query history. In this section, we evaluate the impact on the defense system for different proportions of "\textit{fake queries}" (${p_{fake}}$). Tab. \ref{tab:2} plots the results. It also can be intuitively understood from Fig. \ref{fig-fake}.  As observed, there is a critical threshold ${p_{fake}}$ for the defender: once ${p_{fake}}$ exceeds this threshold, the $DSR$ drops dramatically. For these attacks, if ${p_{fake}} \ge 60\% $, $DSR$ drops to approximately $10\%$ or $0\%$. For CS and DW attacks, the $DSR$ linearly dropped when ${p_{fake}} \in [25,50]$. However, for the other two attacks, this situation does not happen. An intuitive explanation of this can be as follows: ${p_{fake}}$ mainly affects the estimation of the query of interest for defense; yet, the priority of our defense is to distinguish the authenticity of the query,  ${p_{fake}}$ tends to have a larger impact on our proposal. 
The adversary's strategy to evade detection would probably be to set up ${p_{fake}}$ to a sufficiently high value (e.g., ${p_{fake}} \ge 60\% $), but this would dramatically raise the cost of the attack and the number of queries. This makes the attacker overwhelmed and they are not sure if they can obtain AEs to attack target models successfully.

 \;\noindent$\bullet$\textit{\textit{A3. Other Adaptive Attacks.}}
EOT is a well-known attack on images \cite{robust}. However, in audio attacks, after testing EOT transformations in audio (including waveform shifting, volume up/down, Pitch Shifting, Frequency Mask \cite{specaugment}, SpecSwap, etc), we found that EOT transformations could play the role of enlarging the datasets but has no significant effect on ASR results. We suspect that this is attributed to the time-series correlation between the before and after of audio data, so some simple transformations cannot impact ASR. Therefore it is difficult to perform an adaptive attack similar to the images.
We evaluated the two likeliest adaptive attacks, and of course, it is also possible to design the other attack according to the details of the defense, but probably without significant impact.

\textbf{Robust Defense} \label{roboust}

In the \textit{random noise} adaptive attack and Fig. 6, we found that the appropriate level of noise could help us build a more robust defense system, so we further studied the subtle relationship. In Tab. \ref{tab-noise-aug}, we set up six different noise levels. The audio belongs to high-fidelity quality audio when $SNR>75$ and the noise is extremely slight. Once the noise gradually rises to $SNR=75$, our defense system can achieve more than $90\%$ detection success rate for all attacks; when the noise rises to $SNR=50$, the detection success rate reaches the maximum (and the average is $93.69\%$). The noise $SNR<25$, the noise has become significant, exceeds the threshold, and the detection success rate drops. So, with the noise  $SNR \in [25,75]$, we can build a more robust defense system and achieve a detection ratio of over $90\%$. Besides, our experiments also proved that the small input noise has a defense effect \cite{defense_noise}.

\section*{Conclusion and Discussion}\label{con_and_limit}
In this work, we analyze adversary behavior during AE generation and detect potential attacks based on the association before and after the query. 
Our focus is on detecting the AE generation process, which provides a novel approach to process-based defense. Our approach achieves average detection success rate of over $90\%$. It is a lightweight framework that is both quick and efficient, able to be closely combined with other defenses to build the foundation for a structured defense system.

However, with more research on attacks, single-step generation attack of AEs is growing, which impose higher requirements to the defense. From another aspect, our scheme increases the attacker's cost of attack, and our scheme will be fooled if the attacker has a large number of resources. Fingerprint fraud techniques can also create vulnerabilities in our approach. In addition, some adversaries may give up their attacks on the target system and turn to attack the defense system, which also warrants our attention.

\section*{Appendix}
\section{Datasets\label{datasets}}
\subsection{Music-sets\label{music-sets} }
We contacted the authors of CommanderSong \cite{7} and Devil's Whisper \cite{9} to consult them on the details about how to design the music-based carries for the adversarial samples (AEs) they used in their experiments, and obtained a copy of the original music dataset they applied.
To evaluate the threshold, we created a music carrier dataset for making AEs based on the obtained original music dataset. We have released the processed dataset and you can get our data from: \url{https://drive.google.com/file/d/1wPVK9S8TyB0aaXqXFKEebYKuKshmBvDc/view.}

The original music dataset is a raw dataset of $100$ songs collected on YouTube, including pop, classical, rock and light music, and ranging across multiple languages, including Korean, English, Japanese, Chinese, Russian and Arabic. The length of each song is about $5$ minutes.

In our experiments, we studied the impact of different audio lengths on AEs and found that different lengths of audio affect the generation of adversarial examples. Overly short audio decreases the success rate of attacks, and too long audio increases the cost of producing AEs. Only properly lengthy audio is a candidate for AEs. We use Word Error Rates  (WER) to research this issue.
\begin{equation}
\label{wer}
WER = 100\% *\frac{{S + D + I}}{N}
\end{equation}
In Eq. \ref{wer}, $S$ represents the number of characters replaced, $D$ represents the number of characters deleted, $I$ represents the number of characters inserted, and $N$ represents the total number of characters.

\begin{figure}[thp]
\centering  
\includegraphics[width=0.45\textwidth]{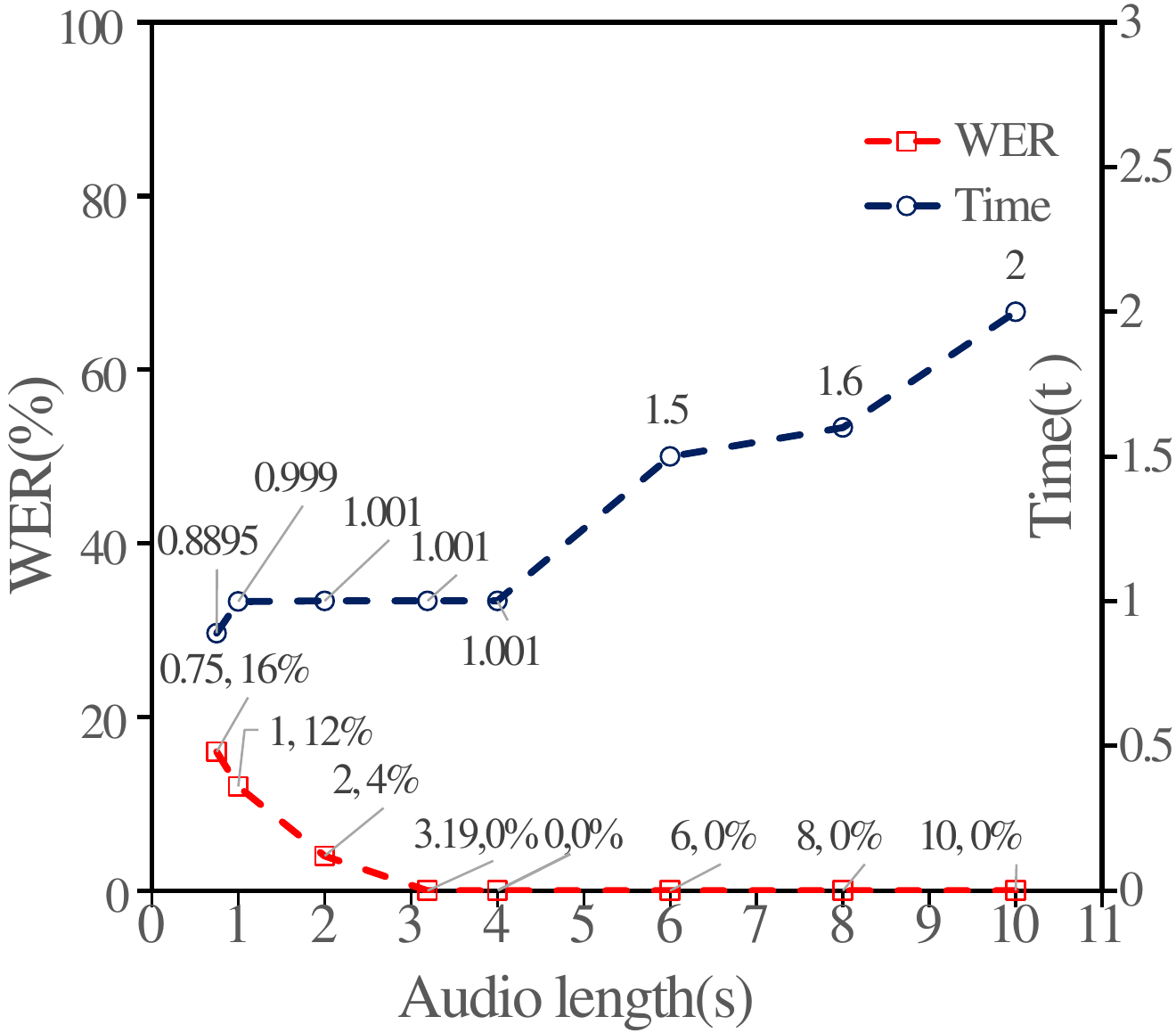}
\caption{\label{length} Audio length impacts the production time of AEs and the integrity of command.}
\end{figure}
 
From Fig. \ref{length}, it can be seen that the WER changes with the length of the audio. If the audio length is below $3.19s$, the attack success rate of the AEs decreases as the audio length reduces (the WER of the target command increases). Above this value, the attack success rate reaches $100\%$ and the WER falls to $0\%$. However, the time cost of producing an AE increases linearly with the length of the audio. The longer the audio, the higher the cost of producing AEs. While the audio length is $3s$-$4s$, the most excellent performance is obtained and the ratio of time cost to WER is the lowest. Finally, the recommended audio length is $3s$ or $4s$ by balancing time and word error rate. During the production of our dataset, we divided each audio data into $3s$ and $4s$ to balance the success rate of the attack and the cost.
 
To simulate disturbances and improve the noise immunity of the audio, we must insert some noise into the clean dataset. Our experiments showed that when music develops as the carrier, the inserted noise is within $8000$ (randomly insert), and the similarity distribution is in $[0.36,1]$. The noise does not influence people's auditory perception, and the primary information of the audio remains reachable. So we keep the randomly inserted noise to the audio below $8000$. When clipping music, the length of each slice is limited to 4s according to the principle of random slice. For each song, segment 25 slices at a time, 5 times in total. Finally, obtaining $5*25*100=12,500$ slices. After that, the noise is randomly inserted into some of these slices by randomly displacing the sequence. After testing each slice, there were $10553$ of qualified slices obtained in total. Storage space occupied nearly 1.3G.

Currently, in the field of audio adversarial attacks, no publicly available dataset is based on music, except for some are dialogue-based which as a carrier for AEs. Instead, music is becoming a necessary candidate for attacks due to some of its advantages, but lack of proper datasets. To alleviate this problem, we are happy to share our data with the researcher community so that they can develop more research on music-based attacks and defenses. We also welcome interested researchers to expand the dataset with us.

\subsection{Mini LibriSpeech \label{mini}}
For the Mini-LibriSpeech dataset, we used FFmpeg\footnote{\url{https://github.com/FFmpeg/FFmpeg}} to convert from flac to wav. According to Fig. \ref{length}, we removed some samples that were either overly short or overly long, and we suggested recalculating the threshold to ensure that the detection was not affected once the dataset was modified. You can download the training data set from \url{https://www.openslr.org/resources/31/train-clean-5.tar.gz}.


\section{Benign examples and AEs Audio Fingerprint}
\label{fingerprint}
\begin{figure*}[htbp]
\centering  
\subfigure[Clean audio of Dialogue-carrier]{
\includegraphics[width=0.225\textwidth]{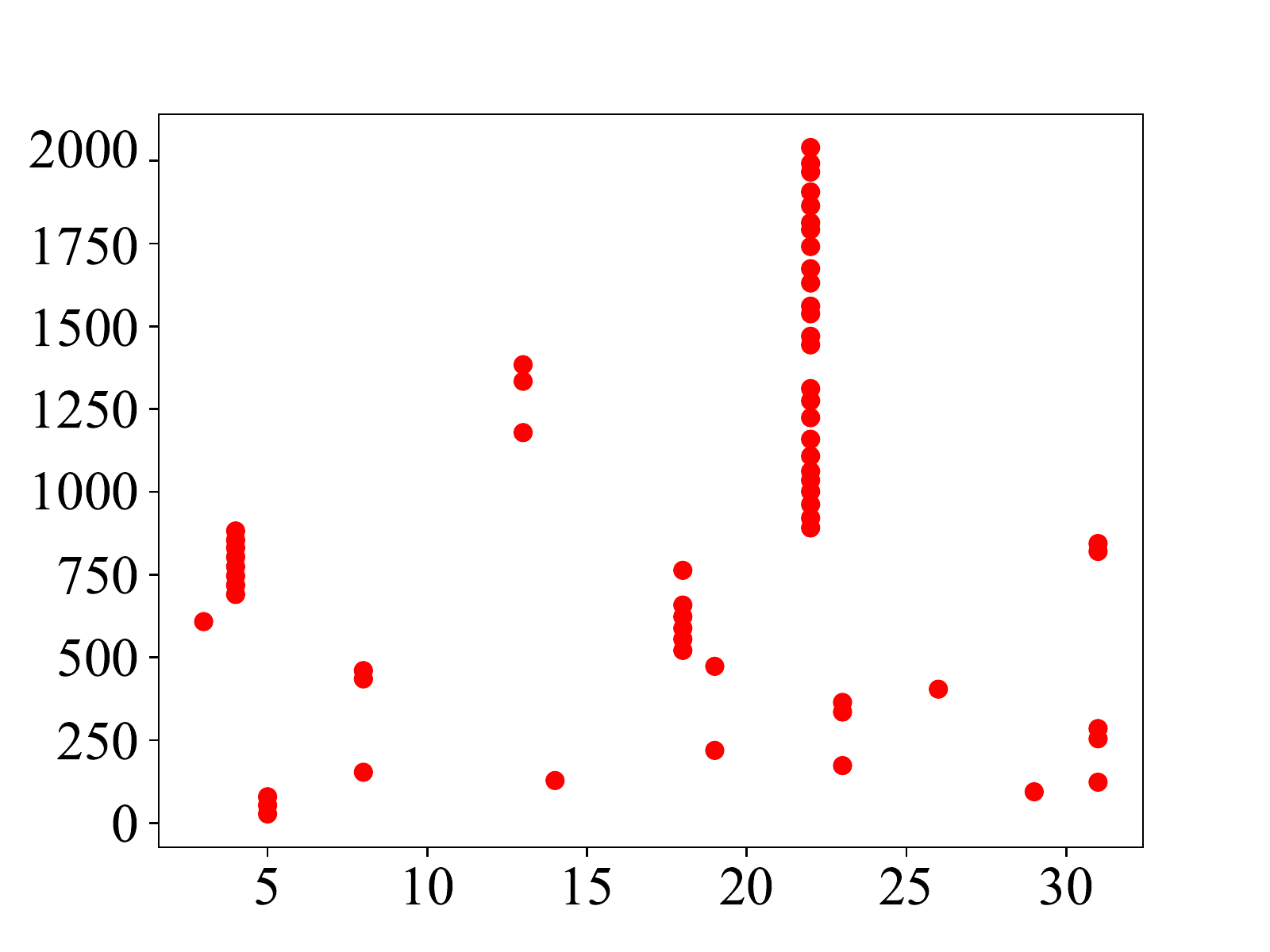}}
\subfigure[AE of Dialogue-carrier ]{
\includegraphics[width=0.225\textwidth]{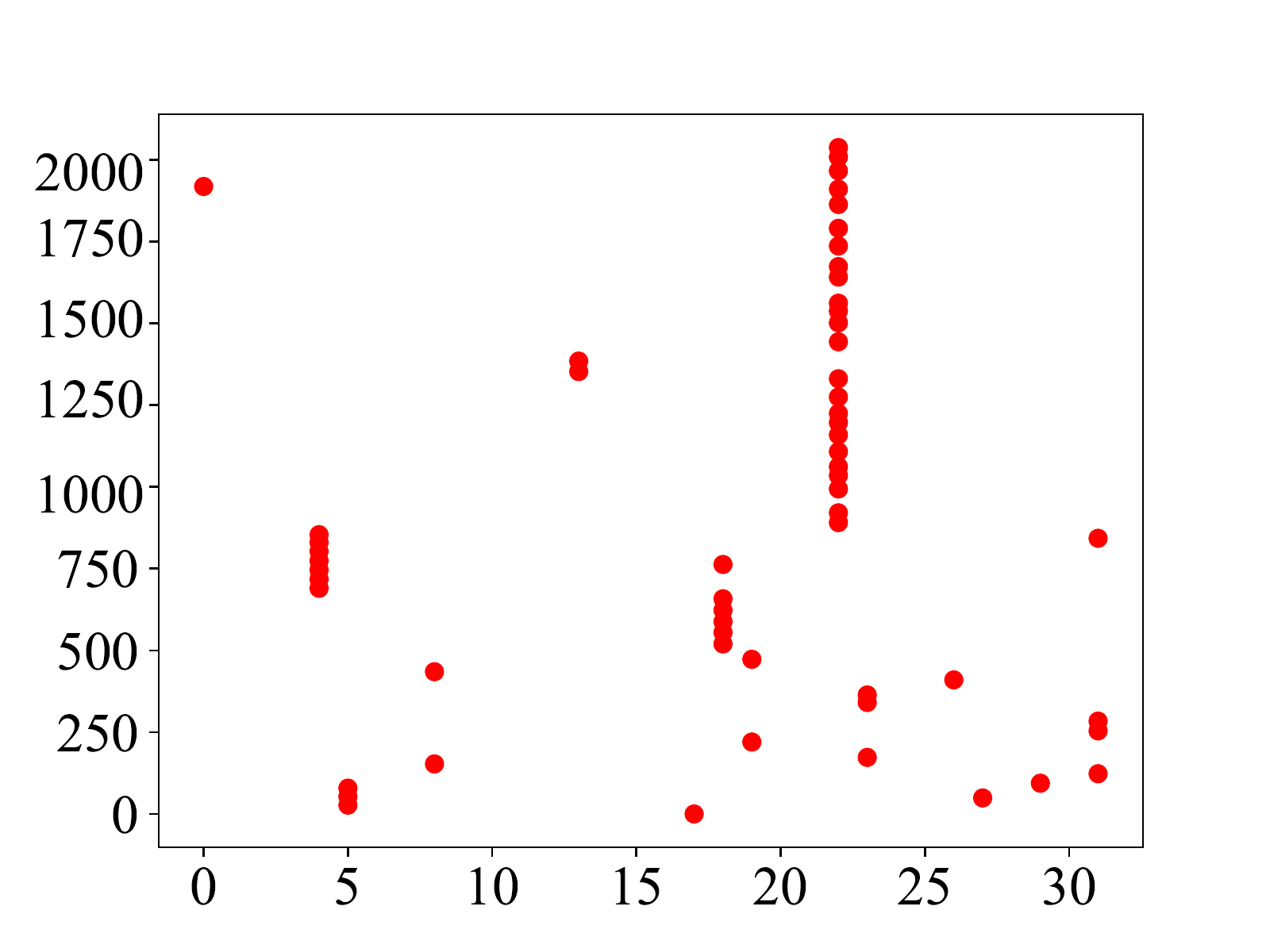}}
\subfigure[Clean audio of Music-carrier]{
\includegraphics[width=0.225\textwidth]{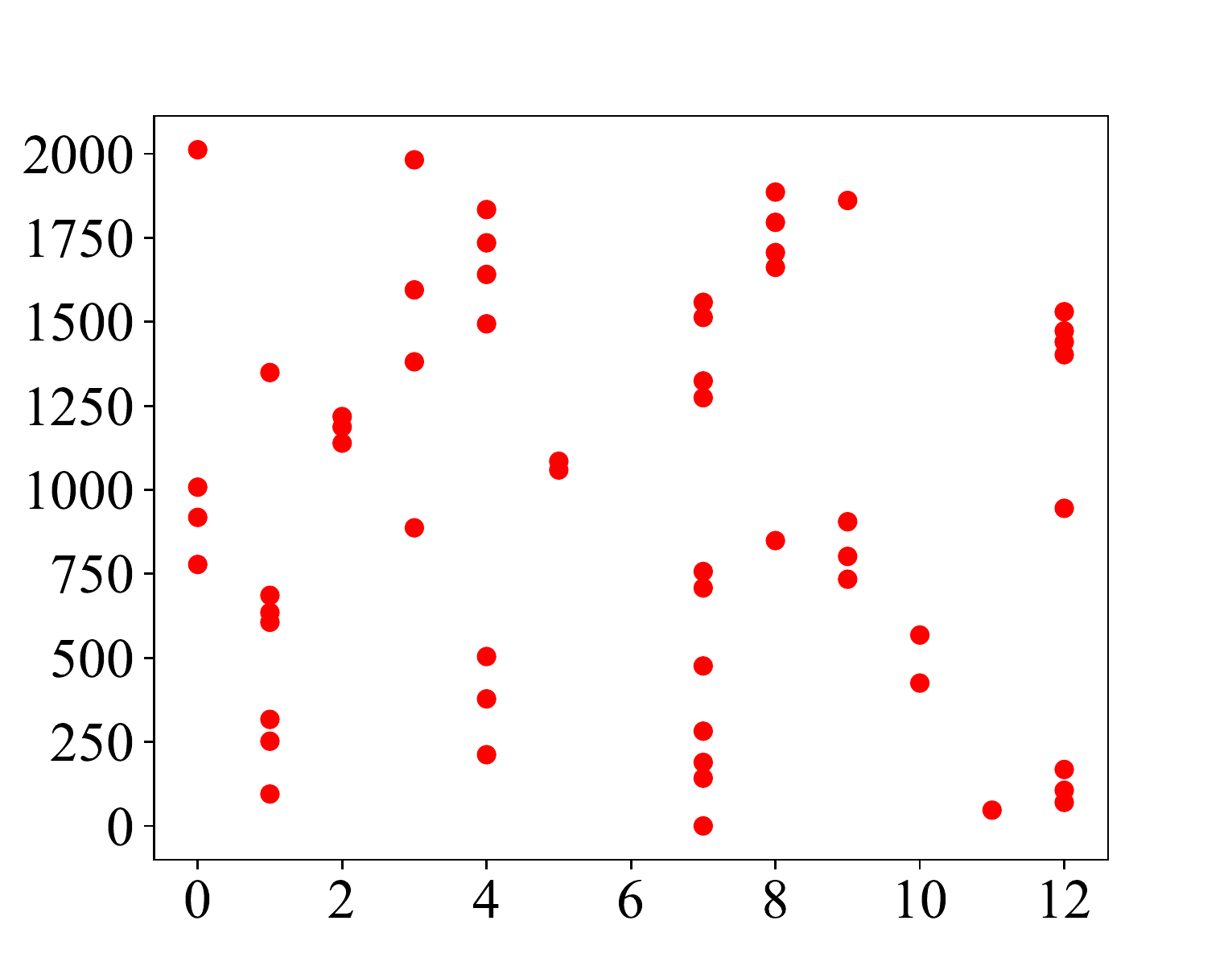}}
\subfigure[AE of Music-carrier]{
\includegraphics[width=0.225\textwidth]{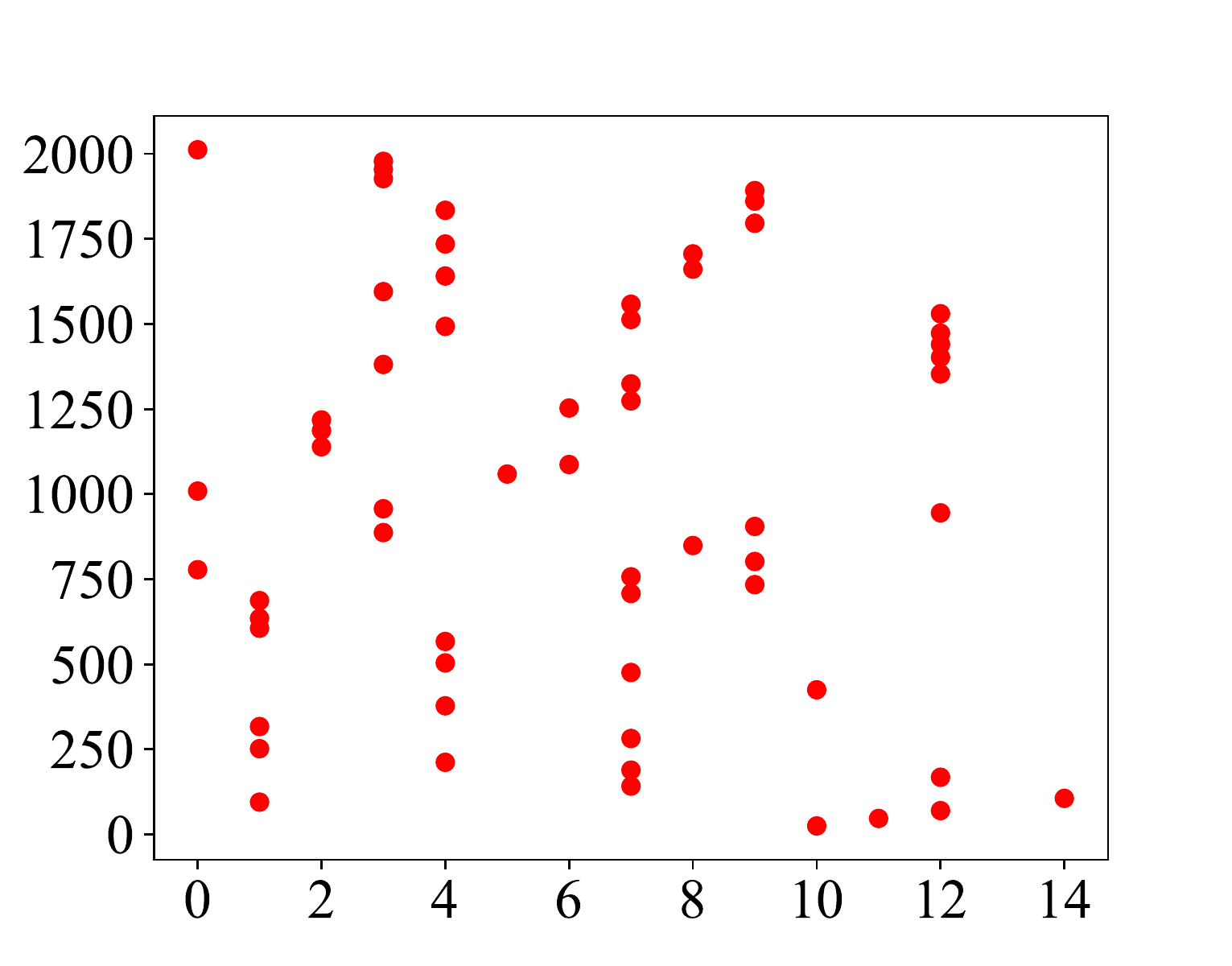}}
\caption{ \label{c-ae-f}  Clean audio and the AEs fingerprint on Dialogue-carrier and Music-carrier.}
\end{figure*}

As shown in Fig. \ref{c-ae-f}, through the addition of perturbations (i.e., noise) on the clean carriers audio to generate AEs, the music-based ones have relatively more and richer fingerprints than the dialogue-based ones, which also confirms that the music-based AEs are easier to detect by our scheme. We also observed that the fingerprint difference between AEs and carriers is small. The fingerprint of each query is similar and the calculated similarity between the queries is very high if the carrier intends to generate AEs. This further proves the viability of our scheme.

\section{ Experimental Environment}
\label{env}
Linux Ubuntu20.0.4 operating system, a 2080Ti GPU with 12G memory, Numpy 1.21.5, Cupy-Cuda 114, 64 CPUs with 256G RAM.

\section{Societal Impacts\label{societal-impacts}}
For the attacks that require querying the ASR model, much of the defense work was mainly concentrated on the processing of inputs to achieve the defense purpose. Only considering the examination of individual inputs, it lost the procedure information and the results are often not reliable. Our scheme, on the other hand, involves considering the totality and continuity of inputs and capturing the neglected information, which can help us better track the adversary behaviors and make an accurate diagnosis. Such a strategy is more consistent with sociology as well. Meanwhile, dialogue-based carriers have lots of limitations in practical applications and it's hard to reproduce in real attack scenarios, which are gradually abandoned by researchers.
Music-based AEs are gradually becoming the mainstream of attacks. The music is easily reproduced in the actual attack scenarios. The danger is very significant if music is hijacked as AEs, which cannot be ignored by researchers. However, the existing evaluation of defense work is still focused on the evaluation of public dialogue datasets. Lack of evaluation of music-based datasets for defense. In our paper, we have comprehensively evaluated the AEs with music-based carriers, which has a large social impact and also lays a solid foundation for related works in the future.


\begin{backmatter}

\section*{Acknowledgements}
Not applicable.

\section*{Funding}
The authors are supported in part by NSFC No.62202275 and Shandong-SF No.ZR2022QF012 projects.

\section*{Abbreviations}
\makebox[1.5cm][l]{AEs}Adversarial Examples \\
\makebox[1.5cm][l]{ASR}Automatic Speech Recognition \\
\makebox[1.5cm][l]{DNNs}Deep Neural Networks \\

\section*{Availability of data and materials}
The code is available at: \url{https://github.com/xx.}


\section*{Competing interests}
The authors declare that they have no competing interests.


\section*{Authors' contributions}



\bibliographystyle{bmc-mathphys} 
\bibliography{refer}      





\end{backmatter}
\end{document}